\documentclass[aps,prl,twocolumn,showpacs,superscriptaddress, floatfix]{revtex4-2}  
\usepackage{graphicx}  
\usepackage{dcolumn}   
\usepackage{bm}        
\usepackage{amssymb}   
\usepackage[caption=false]{subfig}
\usepackage{xcolor}

\makeatletter
\newif\ifpreprintoption
\@ifclasswith{revtex4-2}{preprint}{\preprintoptiontrue}{\preprintoptionfalse}
\makeatother

\newcommand{\minerva}{MINERvA}
\newcommand{\qelike}{quasielastic-like}
\newcommand{\tune}{\minerva~tune v4.4.1}

\newcommand{\pt}{\ensuremath{p_{t}}}
\newcommand{\pz}{\ensuremath{p_{||}}}
\newcommand{\eavail}{\ensuremath{E_{\rm\scriptstyle available}}}
\newcommand{\sumtp}{\ensuremath{\Sigma T_{p}}}

\newcommand{\qzeroqe}{\ensuremath{q_{0}^{\rm\scriptstyle (QE)}}}

\begin{document}




\newcommand{\Rutgers}{Rutgers, The State University of New Jersey, Piscataway, New Jersey 08854, USA}
\newcommand{\Hampton}{Hampton University, Dept. of Physics, Hampton, VA 23668, USA}
\newcommand{\Dortmund}{Institute of Physics, Dortmund University, 44221, Germany }
\newcommand{\Otterbein}{Department of Physics, Otterbein University, 1 South Grove Street, Westerville, OH, 43081 USA}
\newcommand{\JMU}{James Madison University, Harrisonburg, Virginia 22807, USA}
\newcommand{\Florida}{University of Florida, Department of Physics, Gainesville, FL 32611}
\newcommand{\UCIrvine}{Department of Physics and Astronomy, University of California, Irvine, Irvine, California 92697-4575, USA}
\newcommand{\CBPF}{Centro Brasileiro de Pesquisas F\'{i}sicas, Rua Dr. Xavier Sigaud 150, Urca, Rio de Janeiro, Rio de Janeiro, 22290-180, Brazil}
\newcommand{\PUCP}{Secci\'{o}n F\'{i}sica, Departamento de Ciencias, Pontificia Universidad Cat\'{o}lica del Per\'{u}, Apartado 1761, Lima, Per\'{u}}
\newcommand{\INRM}{Institute for Nuclear Research of the Russian Academy of Sciences, 117312 Moscow, Russia}
\newcommand{\Jlab}{Jefferson Lab, 12000 Jefferson Avenue, Newport News, VA 23606, USA}
\newcommand{\Pittsburgh}{Department of Physics and Astronomy, University of Pittsburgh, Pittsburgh, Pennsylvania 15260, USA}
\newcommand{\Guanajuato}{Campus Le\'{o}n y Campus Guanajuato, Universidad de Guanajuato, Lascurain de Retana No. 5, Colonia Centro, Guanajuato 36000, Guanajuato M\'{e}xico.}
\newcommand{\Athens}{Department of Physics, University of Athens, GR-15771 Athens, Greece}
\newcommand{\Tufts}{Physics Department, Tufts University, Medford, Massachusetts 02155, USA}
\newcommand{\WM}{Department of Physics, William \& Mary, Williamsburg, Virginia 23187, USA}
\newcommand{\FNAL}{Fermi National Accelerator Laboratory, Batavia, Illinois 60510, USA}
\newcommand{\Purdue}{Department of Chemistry and Physics, Purdue University Calumet, Hammond, Indiana 46323, USA}
\newcommand{\MCLA}{Massachusetts College of Liberal Arts, 375 Church Street, North Adams, MA 01247}
\newcommand{\UMD}{Department of Physics, University of Minnesota -- Duluth, Duluth, Minnesota 55812, USA}
\newcommand{\Northwestern}{Northwestern University, Evanston, Illinois 60208}
\newcommand{\UNI}{Facultad de Ciencias, Universidad Nacional de Ingenier\'{i}a, Apartado 31139, Lima, Per\'{u}}
\newcommand{\Rochester}{Department of Physics and Astronomy, University of Rochester, Rochester, New York 14627 USA}
\newcommand{\Austin}{Department of Physics, University of Texas, 1 University Station, Austin, Texas 78712, USA}
\newcommand{\USM}{Departamento de F\'{i}sica, Universidad T\'{e}cnica Federico Santa Mar\'{i}a, Avenida Espa\~{n}a 1680 Casilla 110-V, Valpara\'{i}so, Chile}
\newcommand{\Geneva}{University of Geneva, 1211 Geneva 4, Switzerland}
\newcommand{\Chicago}{Enrico Fermi Institute, University of Chicago, Chicago, IL 60637 USA}
\newcommand{\hired}{}
\newcommand{\OregonState}{Department of Physics, Oregon State University, Corvallis, Oregon 97331, USA}
\newcommand{\oxford}{Oxford University, Department of Physics, Oxford, OX1 3PJ United Kingdom}
\newcommand{\umiss}{University of Mississippi, Oxford, Mississippi 38677, USA}
\newcommand{\upenn}{Department of Physics and Astronomy, University of Pennsylvania, Philadelphia, PA 19104}
\newcommand{\AMU}{AMU Campus, Aligarh, Uttar Pradesh 202001, India}
\newcommand{\wroclaw}{University of Wroclaw, plac Uniwersytecki 1, 50-137 Wroa\l{}aw, Poland}
\newcommand{\Mohali}{Department of Physical Sciences, IISER Mohali, Knowledge City, SAS Nagar, Mohali - 140306, Punjab, India}
\newcommand{\CINVESTAV}{Departamento de Fisica Col. San Pedro Zacatenco, 07360 Mexico, DF, Av. Instituto PolitÃ©cnico Nacional, Mexico}
\newcommand{\york}{York University, Department of Physics and Astronomy, Toronto, Ontario, M3J 1P3 Canada}
\newcommand{\ND}{Department of Physics, University of Notre Dame, Notre Dame, Indiana 46556, USA}
\newcommand{\ICL}{The Blackett Laboratory,  Imperial College London,  London SW7 2BW, United Kingdom}
\newcommand{\warwick}{Department of Physics, University of Warwick, Coventry, CV4 7AL, UK}

\newcommand{\mascencioThanks}{Now at Iowa State University, Ames, IA 50011, USA}
\newcommand{\amitbashyalThanks}{Now at  High Energy Physics/Center for Computational Excellence Department, Argonne National Lab, 9700 S Cass Ave, Lemont, IL 60439}
\newcommand{\ricfregianThanks}{now at Department of Physics and Astronomy, University of California at Davis, Davis, CA 95616, USA}
\newcommand{\mateusfcarneiroThanks}{Now at Brookhaven National Laboratory, Upton, New York 11973-5000, USA}
\newcommand{\finerThanks}{Now at Los Alamos National Laboratory, Los Alamos, New Mexico 87545, USA}
\newcommand{\kleykampThanks}{Now at Department of Physics and Astronomy, University of Mississippi, Oxford, MS 38677}
\newcommand{\bamThanks}{Now at University of Minnesota, Minneapolis, Minnesota 55455, USA}
\newcommand{\byaeggyThanks}{Now at Department of Physics, University of Cincinnati,  Cincinnati, Ohio 45221, USA}

\title{Simultaneous measurement of proton and lepton kinematics in
quasielastic-like $\nu_{\mu}$-hydrocarbon interactions from 2 to 20 GeV}

\author{D.~Ruterbories}                   \affiliation{\Rochester}
\author{S.~Akhter}                        \affiliation{\AMU}
\author{Z.~~Ahmad~Dar}                    \affiliation{\WM}  \affiliation{\AMU}
\author{F.~Akbar}                         \affiliation{\AMU}
\author{V.~Ansari}                        \affiliation{\AMU}
\author{M.~V.~Ascencio}\thanks{\mascencioThanks}  \affiliation{\PUCP}
\author{M.~Sajjad~Athar}                  \affiliation{\AMU}
\author{A.~Bashyal}\thanks{\amitbashyalThanks}  \affiliation{\OregonState}
\author{A.~Bercellie}                     \affiliation{\Rochester}
\author{M.~Betancourt}                    \affiliation{\FNAL}
\author{A.~Bodek}                         \affiliation{\Rochester}
\author{J.~L.~Bonilla}                    \affiliation{\Guanajuato}
\author{A.~Bravar}                        \affiliation{\Geneva}
\author{H.~Budd}                          \affiliation{\Rochester}
\author{G.~Caceres}\thanks{\ricfregianThanks}  \affiliation{\CBPF}
\author{T.~Cai}                           \affiliation{\Rochester}
\author{M.F.~Carneiro}\thanks{\mateusfcarneiroThanks}  \affiliation{\OregonState}  \affiliation{\CBPF}
\author{G.A.~D\'{i}az~}                   \affiliation{\Rochester}
\author{H.~da~Motta}                      \affiliation{\CBPF}
\author{J.~Felix}                         \affiliation{\Guanajuato}
\author{L.~Fields}                        \affiliation{\ND}
\author{A.~Filkins}                       \affiliation{\WM}
\author{R.~Fine}\thanks{\finerThanks}     \affiliation{\Rochester}
\author{A.M.~Gago}                        \affiliation{\PUCP}
\author{H.~Gallagher}                     \affiliation{\Tufts}
\author{P.K.Gaur}                         \affiliation{\AMU}
\author{A.~Ghosh}                         \affiliation{\USM}  \affiliation{\CBPF}
\author{S.M.~Gilligan}                    \affiliation{\OregonState}
\author{R.~Gran}                          \affiliation{\UMD}
\author{E.~Haase}                         \affiliation{\UMD}
\author{D.A.~Harris}                      \affiliation{\york}  \affiliation{\FNAL}
\author{S.~Henry}                         \affiliation{\Rochester}
\author{K.~Jacobsen}                      \affiliation{\UMD}
\author{D.~Jena}                          \affiliation{\FNAL}
\author{S.~Jena}                          \affiliation{\Mohali}
\author{J.~Kleykamp}\thanks{\kleykampThanks}  \affiliation{\Rochester}
\author{A.~Klustov\'{a}}                  \affiliation{\ICL}
\author{M.~Kordosky}                      \affiliation{\WM}
\author{D.~Last}                          \affiliation{\upenn}
\author{A.~Lozano}                        \affiliation{\CBPF}
\author{X.-G.~Lu}                         \affiliation{\warwick}  \affiliation{\oxford}
\author{E.~Maher}                         \affiliation{\MCLA}
\author{S.~Manly}                         \affiliation{\Rochester}
\author{W.A.~Mann}                        \affiliation{\Tufts}
\author{C.~Mauger}                        \affiliation{\upenn}
\author{K.S.~McFarland}                   \affiliation{\Rochester}
\author{A.M.~McGowan}                     \affiliation{\Rochester}
\author{B.~Messerly}\thanks{\bamThanks}   \affiliation{\Pittsburgh}
\author{J.~Miller}                        \affiliation{\USM}
\author{O.~Moreno}                        \affiliation{\WM}  \affiliation{\Guanajuato}
\author{J.G.~Morf\'{i}n}                  \affiliation{\FNAL}
\author{D.~Naples}                        \affiliation{\Pittsburgh}
\author{J.K.~Nelson}                      \affiliation{\WM}
\author{C.~Nguyen}                        \affiliation{\Florida}
\author{A.~Olivier}                       \affiliation{\Rochester}
\author{V.~Paolone}                       \affiliation{\Pittsburgh}
\author{G.N.~Perdue}                      \affiliation{\FNAL}  \affiliation{\Rochester}
\author{K.-J.~Plows}                      \affiliation{\oxford}
\author{M.A.~Ram\'{i}rez}                 \affiliation{\upenn}  \affiliation{\Guanajuato}
\author{R.D.~Ransome}                     \affiliation{\Rutgers}
\author{H.~Ray}                           \affiliation{\Florida}
\author{H.~Schellman}                     \affiliation{\OregonState}
\author{C.J.~Solano~Salinas}              \affiliation{\UNI}
\author{H.~Su}                            \affiliation{\Pittsburgh}
\author{M.~Sultana}                       \affiliation{\Rochester}
\author{V.S.~Syrotenko}                   \affiliation{\Tufts}
\author{E.~Valencia}                      \affiliation{\WM}  \affiliation{\Guanajuato}
\author{N.H.~Vaughan}                     \affiliation{\OregonState}
\author{A.V.~Waldron}                     \affiliation{\ICL}
\author{M.O.~Wascko}                      \affiliation{\ICL}
\author{C.~Wret}                          \affiliation{\Rochester}
\author{B.~Yaeggy}\thanks{\byaeggyThanks}  \affiliation{\USM}
\author{L.~Zazueta}                       \affiliation{\WM}



\collaboration{The \minerva\ Collaboration}\ \noaffiliation
\date{\today}

\begin{abstract}
Neutrino charged-current quasielastic-like scattering, a reaction category extensively used in neutrino oscillation measurements, probes nuclear effects that govern neutrino-nucleus interactions.  This Letter reports the first measurement of the triple-differential cross section for $\nu_{\mu}$ quasielastic-like
reactions using the hydrocarbon medium of the MINERvA detector exposed to a wide-band beam spanning 2 $\leq$ E$_\nu \leq$ 20 GeV.  The measurement maps the correlations among transverse and longitudinal muon momenta and summed proton kinetic energies, and compares them to predictions from a state-of-art simulation.  Discrepancies are observed that likely reflect shortfalls with modeling of pion and nucleon intranuclear scattering and/or spectator nucleon ejection from struck nuclei.  The separate determination of leptonic and hadronic variables can inform experimental approaches to neutrino-energy estimation.

\end{abstract}
\maketitle

Current and future long baseline neutrino experiments~\cite{Abe:2015awa,Adamson:2016tbq,Acciarri:2015uup,Abe:2011ts} seek to delineate the neutrino mass ordering and to quantify the presence of  CP violation in the neutrino sector.  These experiments will utilize neutrinos of energies from 0.3 to 4~GeV and higher if tau-neutrino appearance is explored~\cite{DeGouvea:2019kea}.
Accurate models of neutrino-nucleus interactions are required to relate the energies of visible final-state particles of events observed in the detectors to the initiating true neutrino energies that underwrite the oscillations of neutrino flavor. A leading contributor to charged-current (CC) neutrino interactions at these energies is the quasielastic-like channel: 
\begin{equation}
\label{signal-channel}
\nu_{\mu}+\mathcal{A}\rightarrow\mu^{-}+\textrm{nucleons} + \mathcal{A'.}
\label{eqn:qelike-reaction}
\end{equation}
\noindent
Charged-current neutrino interactions within nuclei, even those with apparent two-body quasielastic final states, are altered by a number of poorly understood effects: The struck nucleons of the initial state are bound and in motion~\cite{Moniz:1971mt,Benhar:1994hw}; short-range multinucleon processes give rise to enhanced reaction rates relative to scattering on free nucleons~\cite{Marteau:1999kt,Martini:2010ex,Nieves:2011pp,Gran:2013kda,Fiorentini:2013ezn,Rodrigues:2015hik}, and hadrons produced in the parent $\nu_{\mu}$ 
interactions with nucleons undergo intranuclear final-state interactions (FSI) within the target nuclei.  While the reaction $\nu_{\mu}+\mathcal{A}\rightarrow\mu^{-}+ p + \mathcal{A'}$,
wherein nearly all final-state energy is visible, is thought to be the main
contributor to the quasielastic-like channel (Eq.~\ref{eqn:qelike-reaction}) a significant number of events may have
energy deposited in undetected neutrons or light nuclear fragments. 
Final states of the latter kind complicate the task of inferring neutrino energy 
from samples of quasielastic-like events.

The reaction of Eq.~\ref{eqn:qelike-reaction} has received repeated experimental scrutiny; however only single- or double-differential cross sections in muon kinematics have been reported, mostly carried out
with $\nu_{\mu}$ of incident energies, E$_{\nu}$, of sub-GeV to few GeV~\cite{Gran:2006jn,MiniBooNE:2010bsu,MINERvA:2013kdn,MINERvA:2014ypj,MINOS:2014axb,T2K:2014hih,T2K:2015ujp,T2K:2016jor,T2K:2017qxv,MINERvA:2017dzh,MINERvA:2018hba,MINERvA:2018hqn,T2K:2018rnz,MINERvA:2019ope,MINERvA:2019gsf,T2K:2020jav,MicroBooNE:2020akw,MicroBooNE:2020fxd}.  This Letter
reports a new measurement of the quasielastic-like channel in which the final-state 
muon transverse (\pt) and longitudinal (\pz) momenta are measured in each event simultaneously
with the total ``available" (calorimetrically visible) recoil energy (\eavail) used in previous analyses of data from MINERvA~\cite{Rodrigues:2015hik,MINERvA:2018nab,MINERvA:2021wjs}.
Since the signal requires final state muon plus nucleons only, \eavail\ is  the sum of the kinetic energies of all protons, denoted \sumtp.

Under the assumption of a stationary target neutron in $\nu_\mu + n (bound) \rightarrow \mu^- + p$, energy transfer also can be inferred to be:
\begin{equation}
    \qzeroqe \equiv\frac{m_p^2-(m_n-E_b)^2-m_\mu^2+2(E_\mu-p_\mu cos\theta_\mu)E_\mu}{2\left[(m_n-E_b)-E_\mu+p_\mu cos\theta_\mu\right]}. \label{eqn:q0qe}
\end{equation}
Here, $m_\mu$, $m_p$, and $m_n$ are the masses of the muon, proton, and neutron, $E_b$ is the average binding energy of $34$~MeV~\cite{Bodek:2018lmc,Katori:2008zz,Moniz:1971mt}, and $E_\mu$, $p_\mu$, and $\theta_\mu$ are the muon energy, momentum and angle with respect to the neutrino beam.
The \qzeroqe\ of Eq.~(\ref{eqn:q0qe}) is the quantity added to the reconstructed muon energy by T2K to estimate neutrino energy of quasielastic-like events, while \sumtp\ is the amount added to the muon energy by NOvA to form its neutrino energy estimator for events with quasielastic-like topologies.  Combined T2K-NOvA analyses 
will be credible to the extent that interaction models correctly predict the relationship between these quantities.
The measurements of this Letter elicit the correlations among the kinematic variables of quasielastic channels, thereby confronting the models with information of a kind that heretofore has not been available.

The analysis utilizes high-statistics samples of $\nu_{\mu}$ CC interactions recorded by the \minerva\ detector~\cite{Aliaga:2013uqz} exposed to the wide-band, medium energy NuMI beam~\cite{Adamson:2015dkw} at Fermi National Accelerator Laboratory.  In the NuMI beam, 120 GeV protons impinging upon a carbon target produce pions and kaons that are subsequently charge-selected, focused by a magnetic horn system, and directed into a pipe where they decay.  The resulting neutrino flux is calculated using a GEANT4 simulation of the beam optics with input from hadronic interaction data relevant to the beam and materials~\cite{Aliaga:2016oaz}. The neutrino flux is constrained by previous measurements of neutrino elastic scattering from atomic electrons,
$\nu e^-\to\nu e^-$~\cite{Valencia:2019mkf}. This constraint reduces the normalization uncertainty from 7.8\% to 3.9\% for muon neutrinos of energies between $2$ and $20$ GeV.
%
The neutrino interactions occur in the central scintillator tracker of the MINERvA spectrometer
which has a mass fraction of 88.5\% carbon, 8.2\% hydrogen, 2.5\% oxygen, and trace amounts of other elements. Primary vertices of selected events are restricted to a central 5.3 ton region.  The spatial resolution of the tracker enables reconstruction of final-state protons and Michel electrons from the $\pi^+\to\mu^+(\nu_\mu) \to e^+ (\nu_e\overline{\nu}_\mu)$ decay chain, as well as the tracks of muons.  The magnetized MINOS near detector, located downstream of \minerva, is used to determine the charge and momenta of exiting muons.
The scintillator tracker and the surrounding sampling calorimeters enable calorimetric measurement of \sumtp\ and of photon showers from $\pi^0\to\gamma\gamma$.  Occasionally, final-state neutrons leave a small amount of energy which is tagged as a photon or included in \sumtp; the reference simulation predicts and corrects for this effect. 
The average \sumtp\ for protons is $\approx 250$~MeV;  neutrons contribute less than 10~MeV of energy in 74\% of events and an average of 85 MeV for the rest. 

\begin{figure*}[tp]
    \centering
    \includegraphics[width=0.9\linewidth]{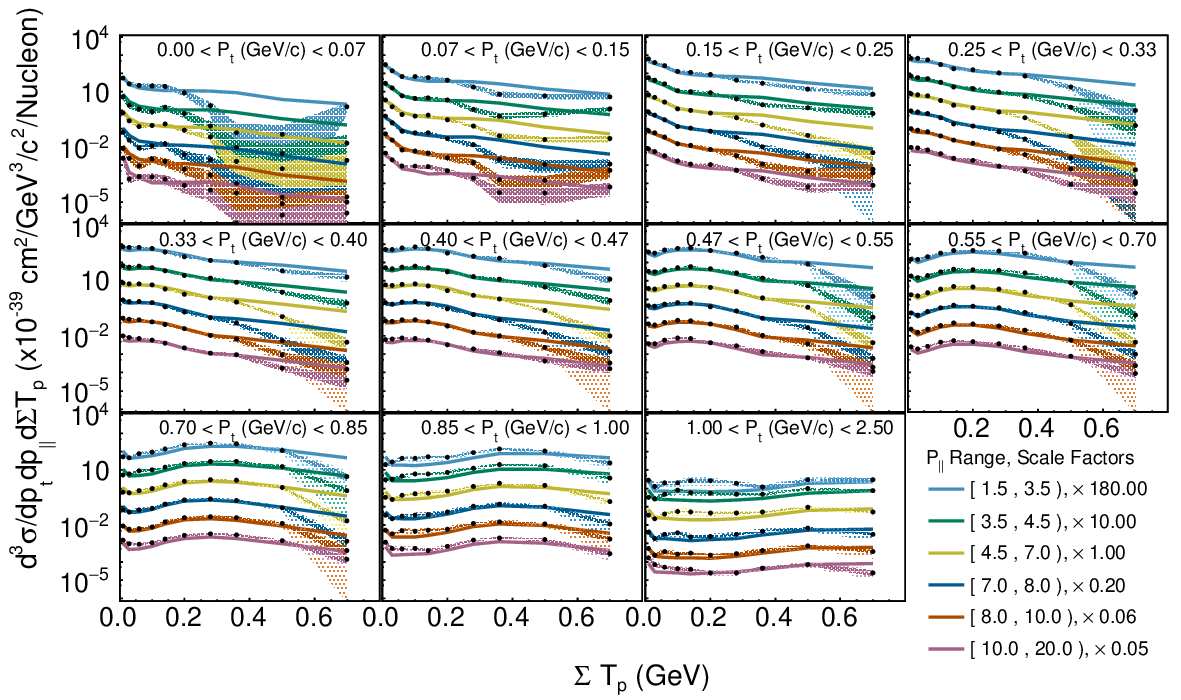}
    \caption{The flux-averaged triple-differential cross section for quasielastic-like events, $d^3\sigma/d\pz d\pt d\sumtp$, shown as points with colored error bands for designated intervals of \pz\ in panels of \pt.   Note the use of scaling factors and log scale to elicit the trends and consistency across all \pz.  The predictions of the reference model \tune\ are shown as lines in each panel.}
    \label{fig:triplexsec}
\end{figure*}
The \minerva ~detector response is simulated using GEANT4~\cite{Agostinelli:2002hh} version 4.9.4p2 with the QGSP\_BERT physics list. The optical and electronics performance is also simulated. Through-going muons are used to determine the absolute energy scale.  Full descriptions of calibrations are given in Refs.~\cite{Aliaga:2013uqz,MINERvA:2021mpk}.
The absolute energy response to charged hadrons is set according to measurements using a charged particle test beam~\cite{Aliaga:2015aqe} and a scaled-down version of the \minerva~detector. The effects of accidental activity as a function of beam intensity are simulated by overlaying hits from data in both \minerva~and MINOS.

The reference signal and background models for this analysis are based on a modified version of the GENIE~\cite{Andreopoulos:2009rq} v.2.12.6 event generator.
Quasielastic interactions are modeled using the Llewellyn Smith formalism~\cite{LlewellynSmith:1971zm} with BBBA05 vector form factors~\cite{Bradford:2006yz} and an axial-vector form factor based on a z-expansion fit to deuterium data~\citep{Meyer:2016oeg}. Resonance production is simulated using the Rein-Sehgal model~\cite{Rein:1980wg} with a dipole axial mass of $M_A^{RES}=1.12$ GeV/c$^2$.  The nuclear initial state is a relativistic Fermi gas model~\cite{Smith:1972xh} with $k_F=0.221$ GeV/c and with a Bodek-Ritchie high momentum tail~\cite{Bodek:1981wr}.  Multinucleon quasielastic-like interactions are simulated by the ``Valencia model" described in Refs.~\cite{Nieves:2011pp,Gran:2013kda,Schwehr:2016pvn}.  Intranuclear final-state interactions of produced hadrons are modeled using the INTRANUKE-hA package~\cite{Dytman:2007zz}.

To better describe \minerva~data, a number of modifications are made in the reference model which are collectively denoted \tune.   The quasielastic cross section is modified as a function of energy and three-momentum transfer based on the random phase approximation (RPA) of the Valencia model~\cite{Nieves:2004wx,Gran:2017psn} appropriate for a Fermi gas~\cite{Martini:2016eec,Nieves:2017lij} to account for long range correlations between nucleons. To account for an observed excess in specific regions of three-momentum transfer and \sumtp , the multinucleon cross section is increased based on fits to \minerva~ data~\cite{Rodrigues:2015hik} from a lower energy beam configuration.  Additionally, based on fits to $\nu_{\mu}$-hydrogen data~\cite{Rodrigues:2016xjj}, the non-resonant CC pion production is decreased by 43\%, the overall baryon-resonance pion production is increased by 15\%, and $M_{A}^{RES}$~is set to 0.94 GeV.

Samples for measuring quasielastic-like interactions and their backgrounds require a muon track that starts in the fiducial volume and is identified in MINOS as negatively charged.  All other tracked particles originating from the interaction vertex at the beginning of the muon track must have $dE/dx$ consistent with a proton.  
Signal and background samples are formed by counting the number of Michel electron candidates within $600$~mm long, $600$~mm diameter cylinders centered on the neutrino vertex and on endpoints of tracked particles, and by counting isolated clusters constructed from two-dimensional clusters with at least 1 MeV visible energy. The former identify $\pi^+$, and the latter identify photons from $\pi^0$ decays. Clusters with an energy less than 10 MeV per hit are assumed to
be caused by neutrons producing low energy protons and are not used.

Events that contain either a $\pi^+$\,(67\%), a $\pi^0$\,(19\%), or both\,(14\%), comprise the dominate backgrounds to the quasielastic-like signal.  Four exclusive samples are assembled using the criteria of $0$ or $\ge 1$ Michel electrons, and $\le 1$ or $\ge 2$ isolated clusters. 
 Sample A with no Michel electrons and $\le 1$ isolated cluster is the signal sample.
Sample B has a Michel electron but $\le 1$ isolated cluster and is rich in single $\pi^+$ events.  Sample C comprises events with $\ge2$ isolated clusters but no Michel electrons, and is mostly single $\pi^0$ events. Sample D events have both Michel electrons and $\ge2$ isolated clusters, and is mostly events with multiple pions.
Details of these four samples are given in Ref.~\cite{MINERvA:2018hqn}.
Sample A has 1.3M selected events with a predicted background of 0.4M.  Samples B-D contain 0.23M, 0.22M, and 57k events.

For each bin of \pt\ and \sumtp, a joint fit to the above-listed four samples is used to determine scale factors applied to the signal sample (A) and to
each of the backgrounds (single $\pi^+$ (B), single $\pi^0$ (C) , and multipion (D)).  The fit minimizes a $\chi^2$ over the four scale factors using a singular value decomposition (SVD) which drops singular values with condition number $<10^{-3}$ to avoid numerical instability and forbids negative scale factors for any component. 
The background-subtracted event rate is unfolded using an iterative technique~\citep{D'Agostini:1994zf} from the RooUnfold framework~\citep{Adye:2011gm} which is regularized by the number of iterations. A regularization of ten iterations was chosen by generating randomly fluctuated pseudodata samples with a number of different underlying physics models to ensure fidelity with different assumed data models.   The statistical covariance matrix is scaled to account for the finite Monte Carlo 
statistics in the true-to-reconstructed migration matrix.
The unfolded 3D distribution is then corrected for the predicted event loss from selection inefficiencies and detector effects. 
The average efficiency is between 40\% and 75\% range over all bins.
The triple-differential cross section is obtained by normalizing the distribution 
according to the number of neutrinos incident on the detector and the number of scattering centers. The final result is shown in Fig.~\ref{fig:triplexsec}.  To zeroth order, Fig.~\ref{fig:triplexsec} shows that the reference simulation (solid lines of different colors in the panels) describes the general trends in the data points.  Upon closer inspection, discrepancies are apparent. 
In the lower \pt\ range of the uppermost panels, for example, the prediction exceeds the data for all \pz\ for \sumtp~$\ge 0.2$\,GeV.

\begin{figure}[th]
    \centering
    \includegraphics[width=0.9\linewidth]{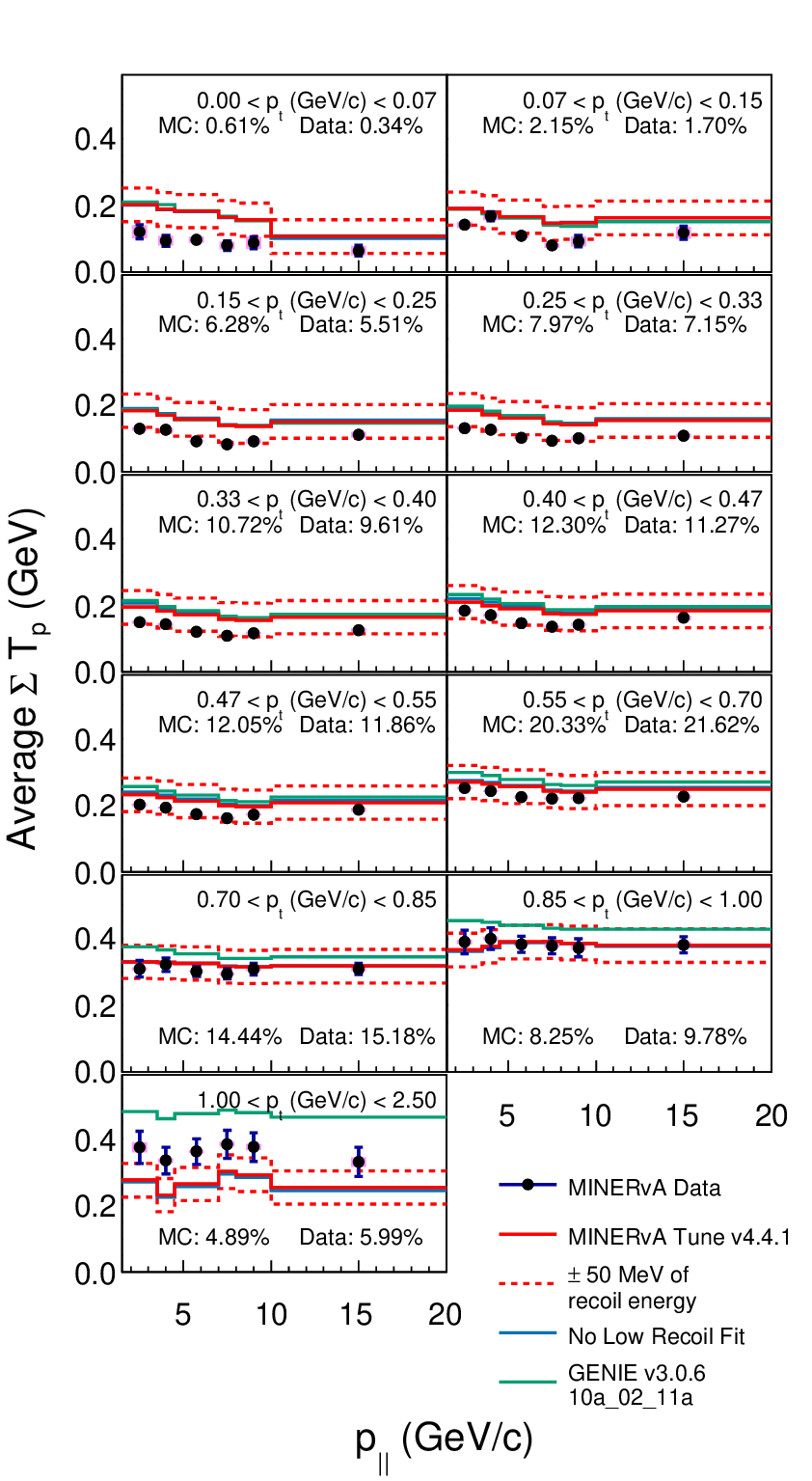}
    \caption{Distribution of average \sumtp~in \pt-\pz~bins. Legends within each panel
    give the percentage contribution of the \pt~panel to the total integrated cross section 
    for data and for the reference simulation. Statistical uncertainty is denoted by the colored box while the total uncertainty by the error bar. 
    The ``No Low Recoil Fit" prediction has no enhancement applied to the 2p2h model.} 
    \label{fig:avgrecoil}
\end{figure}

Figure~\ref{fig:avgrecoil} shows the average \sumtp\ in each $\pt-\pz$ bin.
The average recoil energy in data falls $\sim 50$~MeV below the reference model at low \pt, $\lesssim 0.5$~GeV/c, then rises to be comparable to the model $\sim 0.9$~GeV/c, and finally exceeds the model prediction in the highest $\pt$ bin.  The abrupt change in the highest $\pt$ bins may be due to a cutoff in the Valencia multinucleon model that eliminates this process above three-momentum transfer of $1.2$~GeV/c~\cite{Nieves:2011pp,Gran:2013kda}. Predictions for GENIE v3.0.6 were produced using NUISANCE \cite{Stowell:2016jfr}.

\begin{figure}[tp]
    \centering
    \includegraphics[width = 0.9\linewidth]{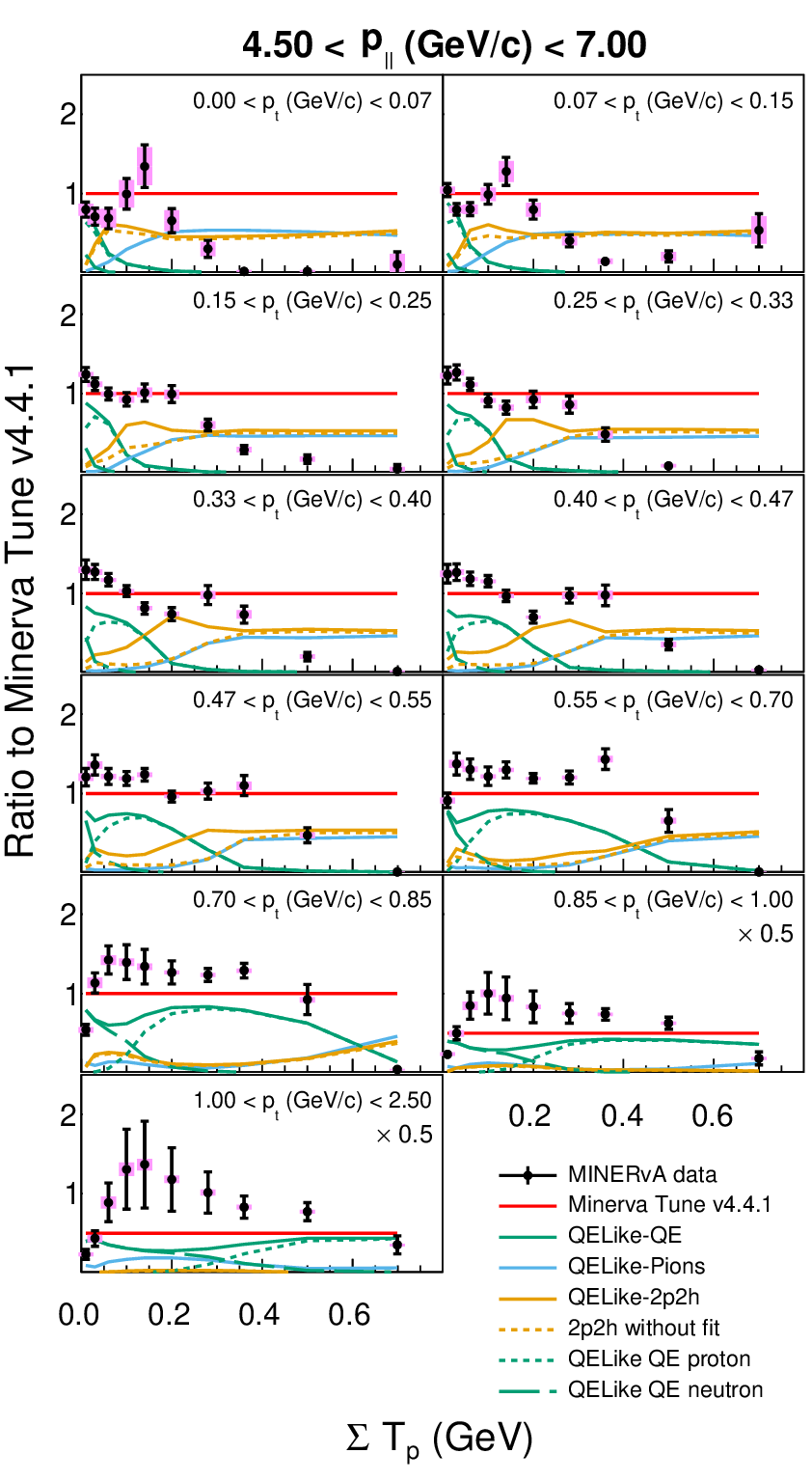}
    \caption{
    Ratio of the differential \qelike~cross-section to predictions in panels of $\pt$ for the peak bin $4.5<\pz<7.0$~GeV/c. The predicted cross section in the reference model is broken down into different contributions. Note that in the highest $\pt$ panels, above $0.85$~GeV/c, the ratio is shown scaled by $0.5$ relative to the other panels. The ``2p2h without fit" is the prediction without the 2p2h-enhancement. 
    The QE subsample is labeled by the identity of the highest energy nucleon as ``QE proton" and ``QE neutron". 
    The ``QELike-Pions" category includes all events with primary pion production followed by FSI absorption of pions.}
    \label{fig:suppression}
\end{figure}

Figures ~\ref{fig:triplexsec} and \ref{fig:avgrecoil} show discrepancies between data and the reference simulation that persist as a function of $\pz$. 
Figure~\ref{fig:suppression} displays the ratio of data to simulation at the peak $\pz$ bin, the third from top (yellow) points in Fig~\ref{fig:triplexsec}. 
Here the discrepancies can be associated with three different kinematic regions of $\pt-\sumtp$, enumerated below.  Since the prominent processes in each region can be identified, each discrepancy suggests specific model developments.

\noindent
{\it Low \pt~, high $\sumtp$:} For $0<\pt<0.4$ and \sumtp$~\ge~$0.3-0.8 GeV, the reference model predicts significant cross section in a region where the data indicates very little cross section. The predicted events are approximately half multinucleon and half low $W$ pion production where the pion is absorbed in FSI. Therefore it is likely that both processes are over-predicted. In this region, the multinucleon prediction is from the Valencia model with little effect from \tune.  In pion production, such an effect may arise either from over-prediction of the baryon-resonance pion production cross sections or from too-small suppression of primary pions due to FSI.
A low $Q^2$ suppression of resonant pion production\cite{Stowell:2019zsh,MINOS:2014axb} or a reduction in the visible energy from these events\cite{MINERvA:2021wjs}, as Fig.~\ref{fig:avgrecoil} also suggests, would improve agreement with this data.
Shifts in energy transfers have been observed in (e,e$'$) data for regions of low energy transfer~\cite{Bodek:2020wbk, Bodek:2018lmc, Ankowski:2014yfa, Boffi:1993gs, Cooper:1993nx, OConnell:1990njm, Sealock:1989nx, Horikawa:1980cg, electronsforneutrinos:2020tbf, Ankowski:2020qbe}. An over-prediction of pion FSI could arise from finite hadronic formation time~\cite{Golan:2012wx}, an effect not included in the reference simulation.  However this background arises mostly from absorption of slow pions (p$_{\pi} < 0.3$\,GeV/c), hence pion formation time is unlikely to account for the entire effect.

\noindent
{\it Moderate \pt~and $\sumtp$~just above the quasielastic peak:} For \sumtp~of 0.2~GeV and $0.15<\pt<0.55$~GeV/c, where the modifications of \tune\ to  multinucleon processes are large,  the data and reference model would be in strong disagreement without these modifications.
Figure~\ref{fig:suppression} shows that the ratio of the data to the reference model dips near the peak of the tune, suggesting that the shape of the \tune\ enhancement may not be accurate, either in rate or in fraction of events with a neutron in the final state.  
However, at $\pt>0.55$~GeV/c, where the model predicts a smaller multinucleon contribution, the data mostly exceeds the reference prediction, suggesting that a significant enhancement to multinucleon processes at higher $\pt$ than in \tune\ may be needed.  

\noindent
{\it High \pt~and low $\sumtp$:} At $\pt>0.55$~GeV/c and $\sumtp<$~50 MeV, there is a significant over-prediction relative to data.  This region is dominated by true quasielastic events where the final-state proton undergoes FSI and leaves the nucleus as one or more energetic neutrons; this suggests that
too much strength is given to FSI in this kinematic region.

\begin{figure}[tp]
    \centering
    \includegraphics[width= 0.9\linewidth]{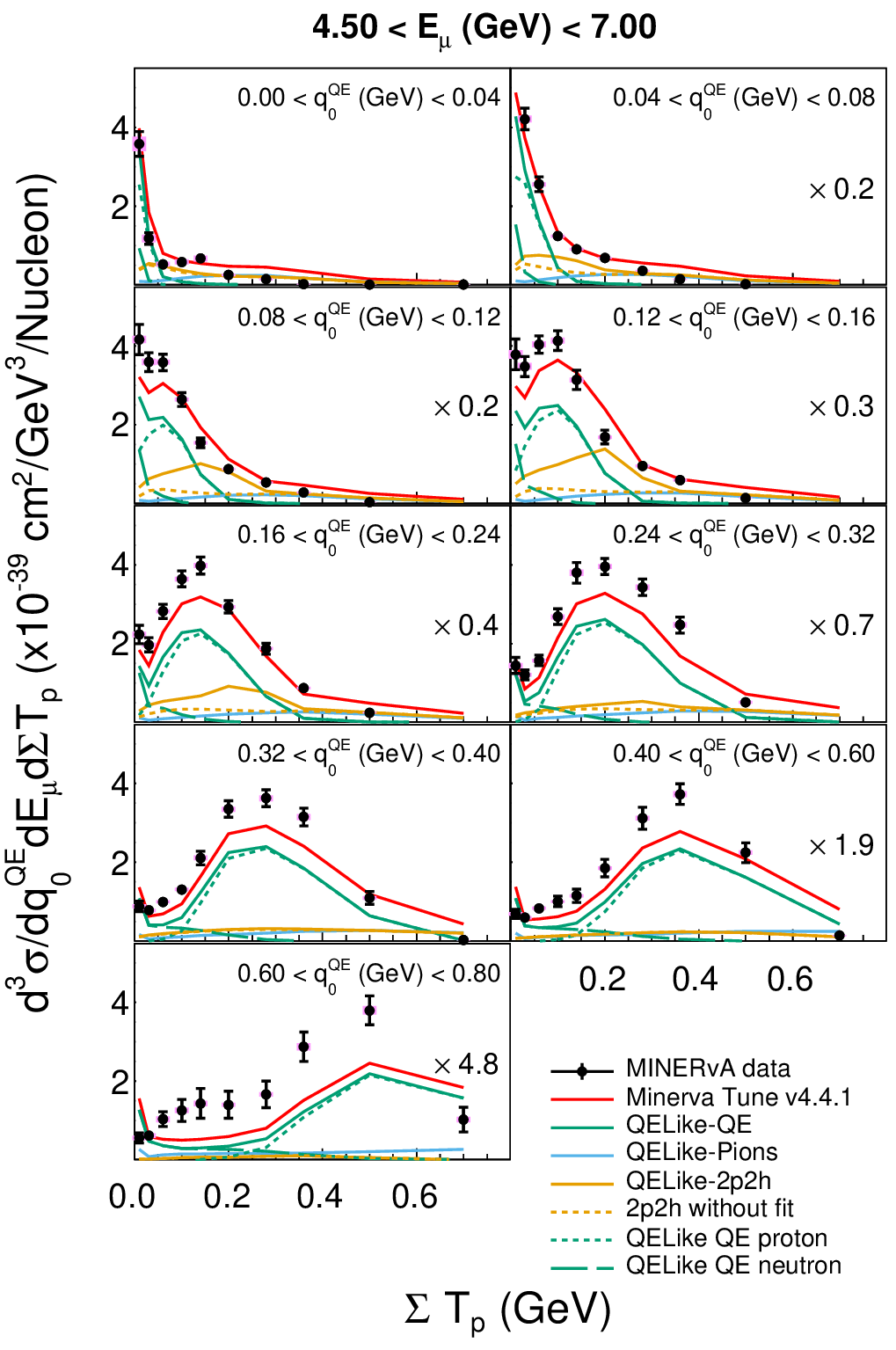}
    \caption{Flux-averaged triple differential cross section, $d^3\sigma/dE_\mu d\qzeroqe d\sumtp$. Note that scale factors are applied to each of the $\qzeroqe$ panels.}
    \label{fig:q0qeehad}
\end{figure}

Figure~\ref{fig:q0qeehad} presents the flux-averaged triple-differential cross section $d^3\sigma/dE_\mu d\qzeroqe d\sumtp$.  Here as well, significant data versus reference model discrepancies are seen at low $\qzeroqe$ for $\sumtp$ beyond the peak of the quasielastic contribution.  The previously noted discrepancy at low $\sumtp$ and high $\pt$ corresponds to a predicted peak near zero $\sumtp$ at high $\qzeroqe$ which is absent from the data.   The cross section in the quasielastic peak is underestimated, especially at higher $\qzeroqe$, and modified form factors could improve this agreement\cite{Meyer:2016oeg,Borah:2020gte,Meyer:2022mix,TejinNature}.  This measurement directly probes the relationship between energy estimators used in oscillation experiments, and discrepancies with models suggest deficiencies in modeling those estimators.

In summary, a number of modeling shortfalls for neutrino-nucleus quasielastic-like scattering are identified by this measurement.  These imply that relationships between the true neutrino energy of quasielastic-like events and experimental estimators, such as $\eavail+E_\mu$ and $\qzeroqe+E_\mu$, differ from those predicted by current neutrino generators.  The triple-differential cross section presented in Fig.~\ref{fig:q0qeehad}  can serve as a benchmark for neutrino-nucleus interaction simulations employed  in ongoing and future neutrino oscillation experiments.

\begin{acknowledgements}
This document was prepared by members of the MINERvA Collaboration using the resources of the Fermi National Accelerator Laboratory (Fermilab), a U.S. Department of Energy, Office of Science, HEP User Facility. Fermilab is managed by Fermi Research Alliance, LLC (FRA), acting under Contract No. DE-AC02-07CH11359.
These resources included support for the MINERvA construction project, and support
for construction also
was granted by the United States National Science Foundation under
Award No. PHY-0619727 and by the University of Rochester. Support for
participating scientists was provided by NSF and DOE (USA); by CAPES
and CNPq (Brazil); by CoNaCyT (Mexico); by  ANID PIA / APOYO AFB180002, CONICYT PIA ACT1413, and Fondecyt 3170845 and 11130133 (Chile); 
by CONCYTEC (Consejo Nacional de Ciencia, Tecnolog\'ia e Innovaci\'on Tecnol\'ogica), DGI-PUCP (Direcci\'on de Gesti\'on de la Investigaci\'on  - Pontificia Universidad Cat\'olica del Peru), and VRI-UNI (Vice-Rectorate for Research of National University of Engineering) (Peru); NCN Opus Grant No. 2016/21/B/ST2/01092 (Poland); by Science and Technology Facilities Council (UK); by EU Horizon 2020 Marie Skłodowska-Curie Action; by an Imperial College London President's PhD Scholarship. D. Ruterbories gratefully acknowledges support from a Cottrell Postdoctoral Fellowship, Research Corporation for Scientific Advancement award number 27467 and National Science Foundation Award CHE2039044.  We thank the MINOS Collaboration for use of its near detector data. Finally, we thank the staff of
Fermilab for support of the beam line, the detector, and computing infrastructure.
\end{acknowledgements}

\bibliography{main}

\end{document}


\begin{center}
\textbf{\large Supplemental Materials}
\end{center}

\section{Presentation of all slices of the Triple Differential Cross Sections}
\begin{figure}[p]
    \centering
    \includegraphics[width = 0.5\linewidth]{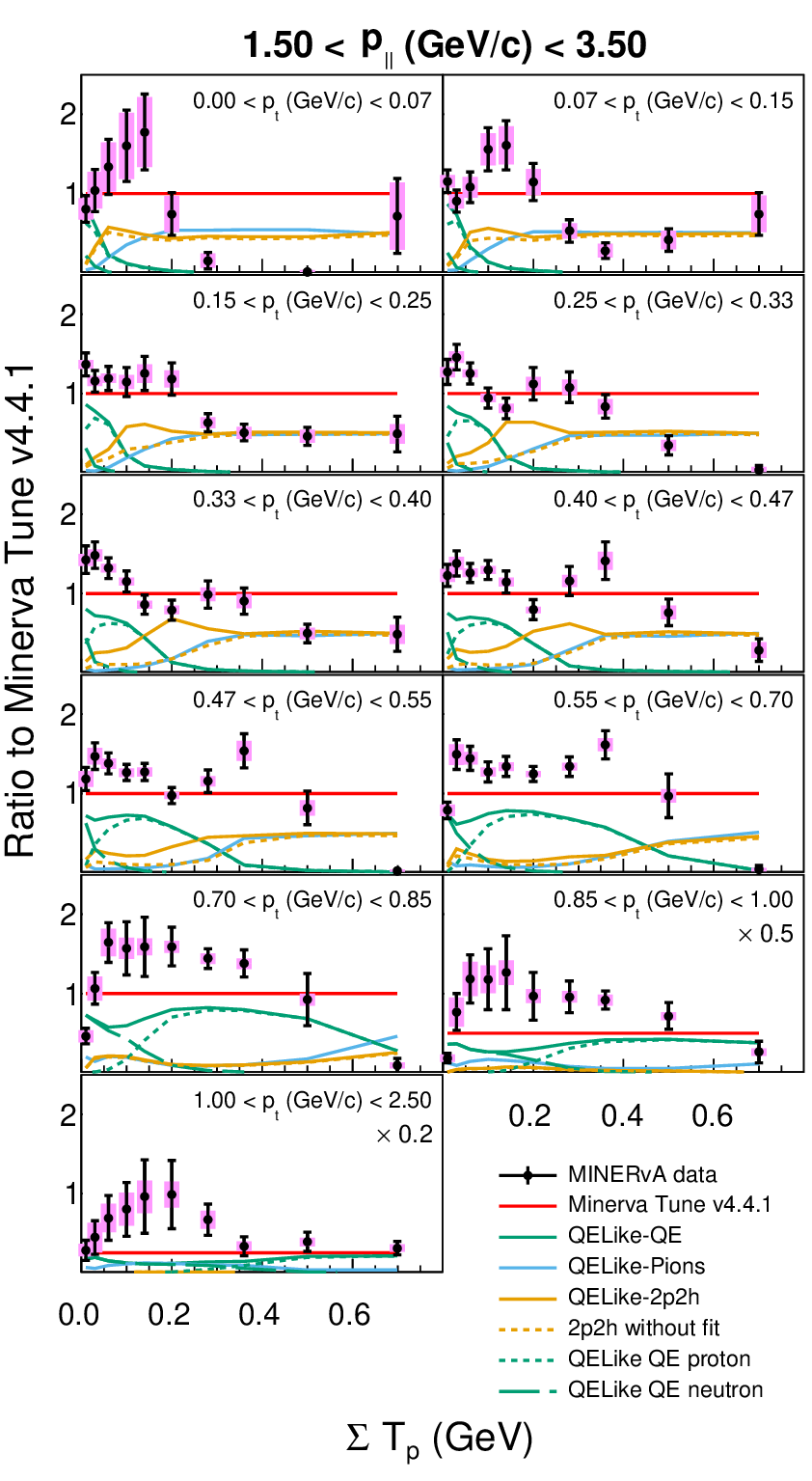}
    \caption{Ratio of the differential cross-section in panels of $\pt$ for the peak bin $1.5<\pz<3.5$~GeV/c.  The predicted cross section in the reference model is broken down into different contributions.  Note that in the highest $\pt$ panels, above $0.85$~GeV/c, the ratio is shown scaled by $0.5$ and $0.2$ relative to the other panels.}
    \label{fig:sup_pzptsumtp_bin1}
\end{figure}

\begin{figure}[tp]
    \centering
    \includegraphics[width = 0.5\linewidth]{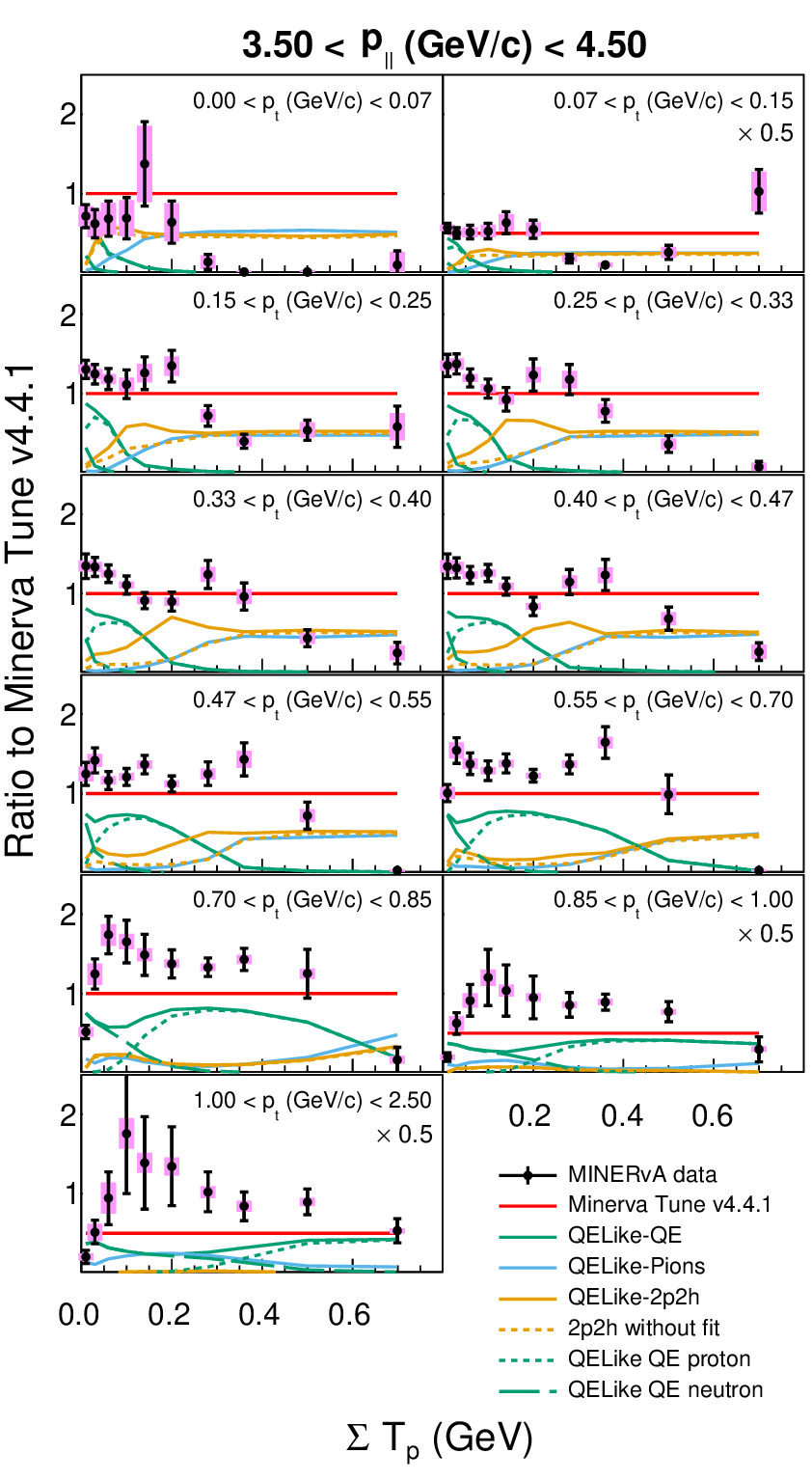}
    \caption{Ratio of the differential cross-section in panels of $\pt$ for the peak bin $3.5<\pz<4.5$~GeV/c.  The predicted cross section in the reference model is broken down into different contributions.  Note that in the highest $\pt$ panels, above $0.85$~GeV/c, the ratio is shown scaled by $0.5$ relative to the other panels.}
    \label{fig:sup_pzptsumtp_bin2}
\end{figure}

\begin{figure}[tp]
    \centering
    \includegraphics[width = 0.5\linewidth]{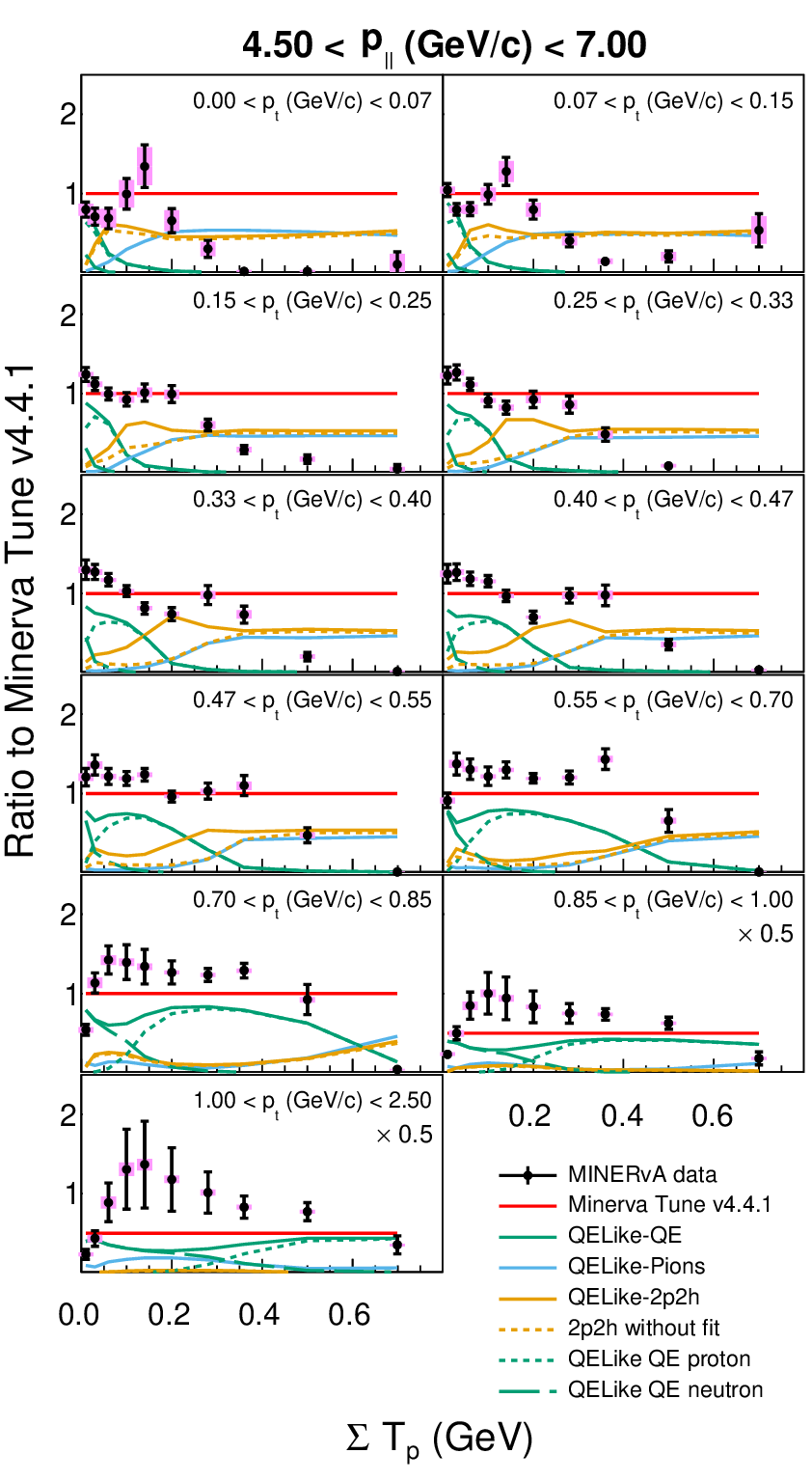}
    \caption{Ratio of the differential cross-section in panels of $\pt$ for the peak bin $4.5<\pz<7.0$~GeV/c.  The predicted cross section in the reference model is broken down into different contributions.  Note that in the highest $\pt$ panels, above $0.85$~GeV/c, the ratio is shown scaled by $0.5$ relative to the other panels.}
    \label{fig:sup_pzptsumtp_bin3}
\end{figure}

\begin{figure}[tp]
    \centering
    \includegraphics[width = 0.5\linewidth]{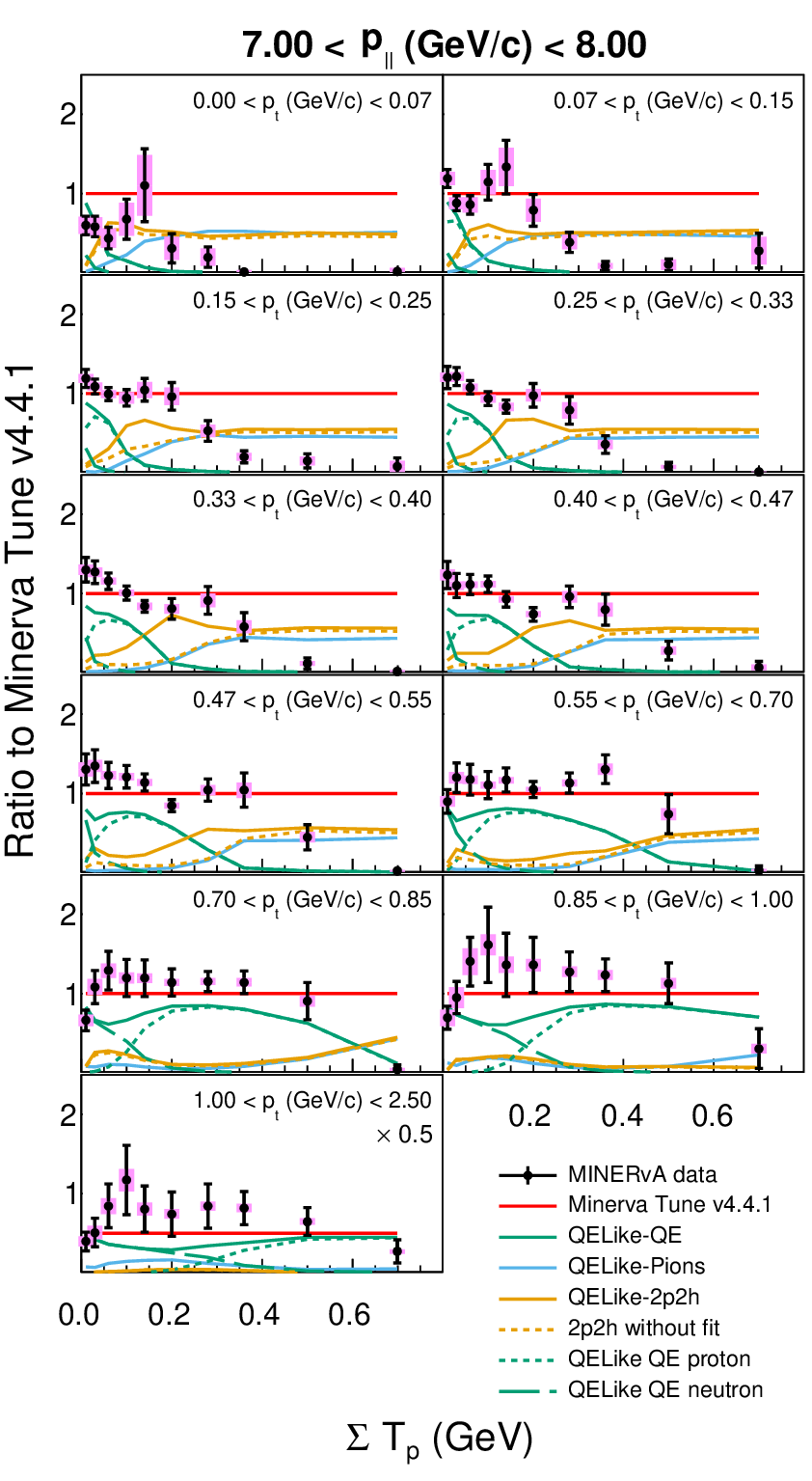}
    \caption{Ratio of the differential cross-section in panels of $\pt$ for the peak bin $7.0<\pz<8.0$~GeV/c.  The predicted cross section in the reference model is broken down into different contributions.  Note that in the highest $\pt$ panels, above $1.0$~GeV/c, the ratio is shown scaled by $0.5$ relative to the other panels.}
    \label{fig:sup_pzptsumtp_bin4}
\end{figure}

\begin{figure}[tp]
    \centering
    \includegraphics[width = 0.5\linewidth]{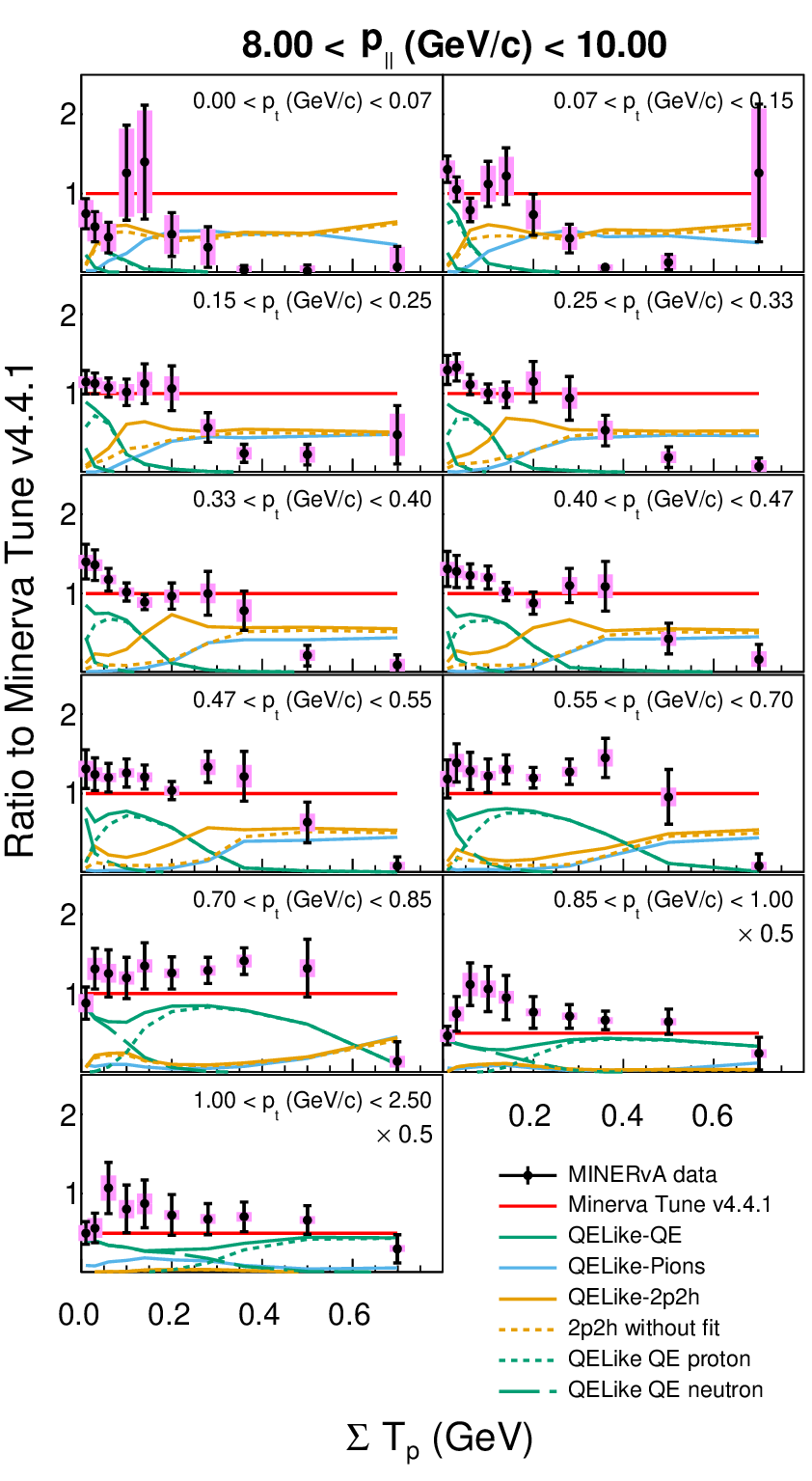}
    \caption{Ratio of the differential cross-section in panels of $\pt$ for the peak bin $8.0<\pz<10.0$~GeV/c.  The predicted cross section in the reference model is broken down into different contributions.  Note that in the highest $\pt$ panels, above $0.85$~GeV/c, the ratio is shown scaled by $0.5$ relative to the other panels.}
    \label{fig:sup_pzptsumtp_bin5}
\end{figure}

\begin{figure}[tp]
    \centering
    \includegraphics[width = 0.5\linewidth]{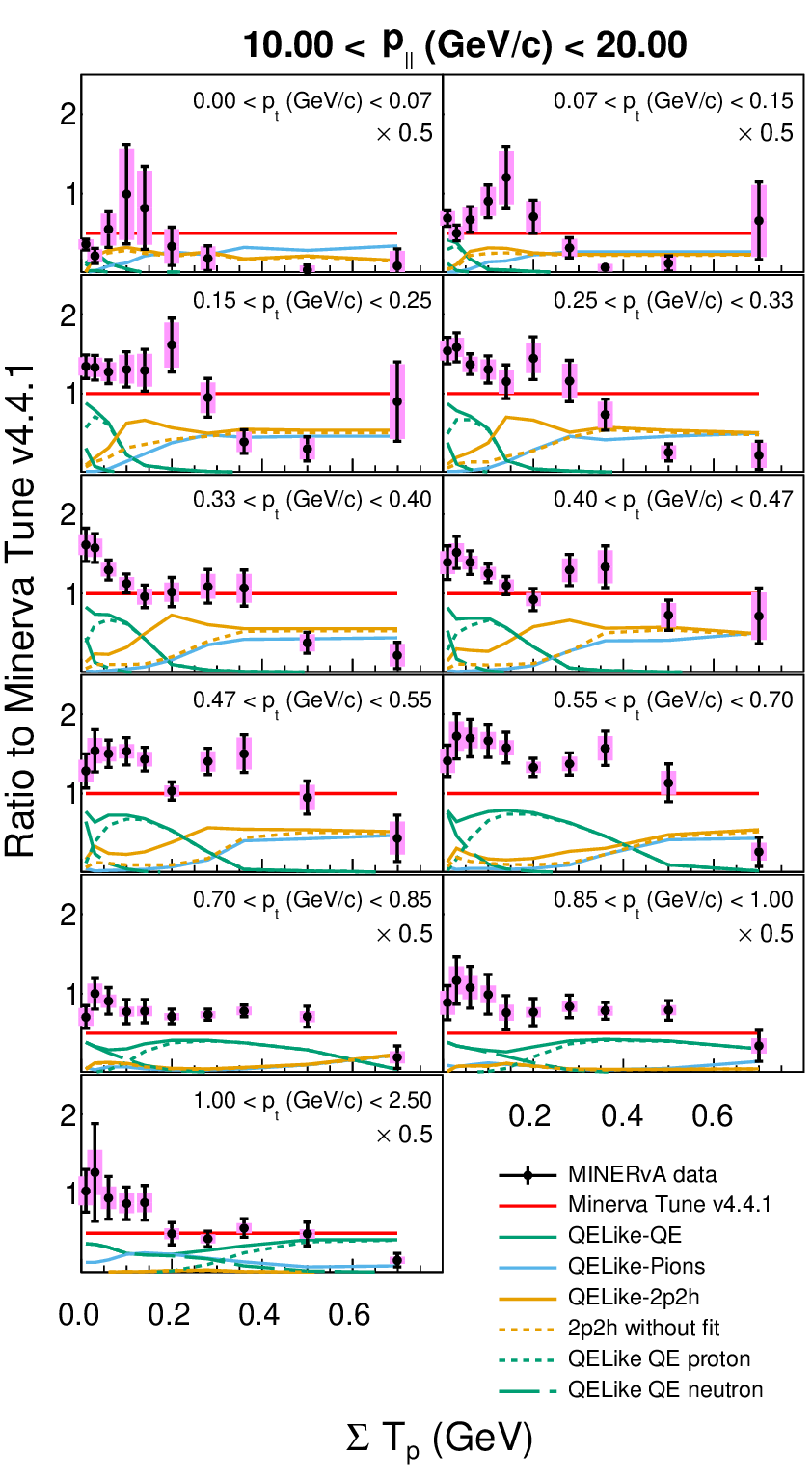}
    \caption{Ratio of the differential cross-section in panels of $\pt$ for the peak bin $10.0<\pz<20.0$~GeV/c.  The predicted cross section in the reference model is broken down into different contributions.  Note that in the highest $\pt$ panels, above $0.70$~GeV/c, the ratio is shown scaled by $0.5$ relative to the other panels.}
\label{fig:sup_pzptsumtp_bin6}
\end{figure}

\begin{figure}[tp]
    \centering
    \includegraphics[width=0.6\linewidth]{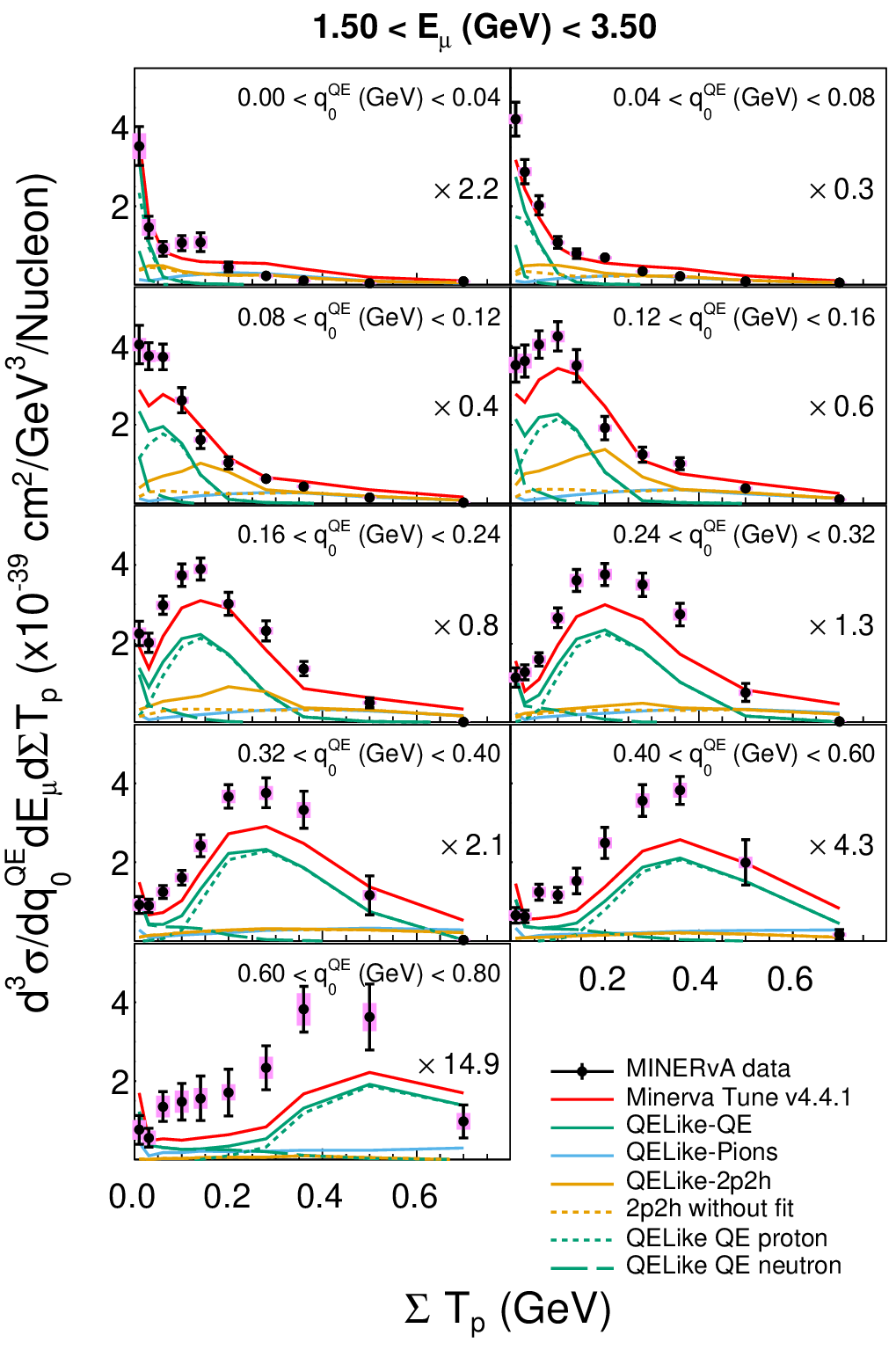}
    \caption{Flux-averaged triple differential cross section, $d^3\sigma/dE_\mu d\qzeroqe d\sumtp$.  Note that scale factors are applied to each of the $\qzeroqe$ panels.}
    \label{fig:sup_q0emusumtp_bin1}
\end{figure}

\begin{figure}[tp]
    \centering
    \includegraphics[width= 0.6\linewidth]{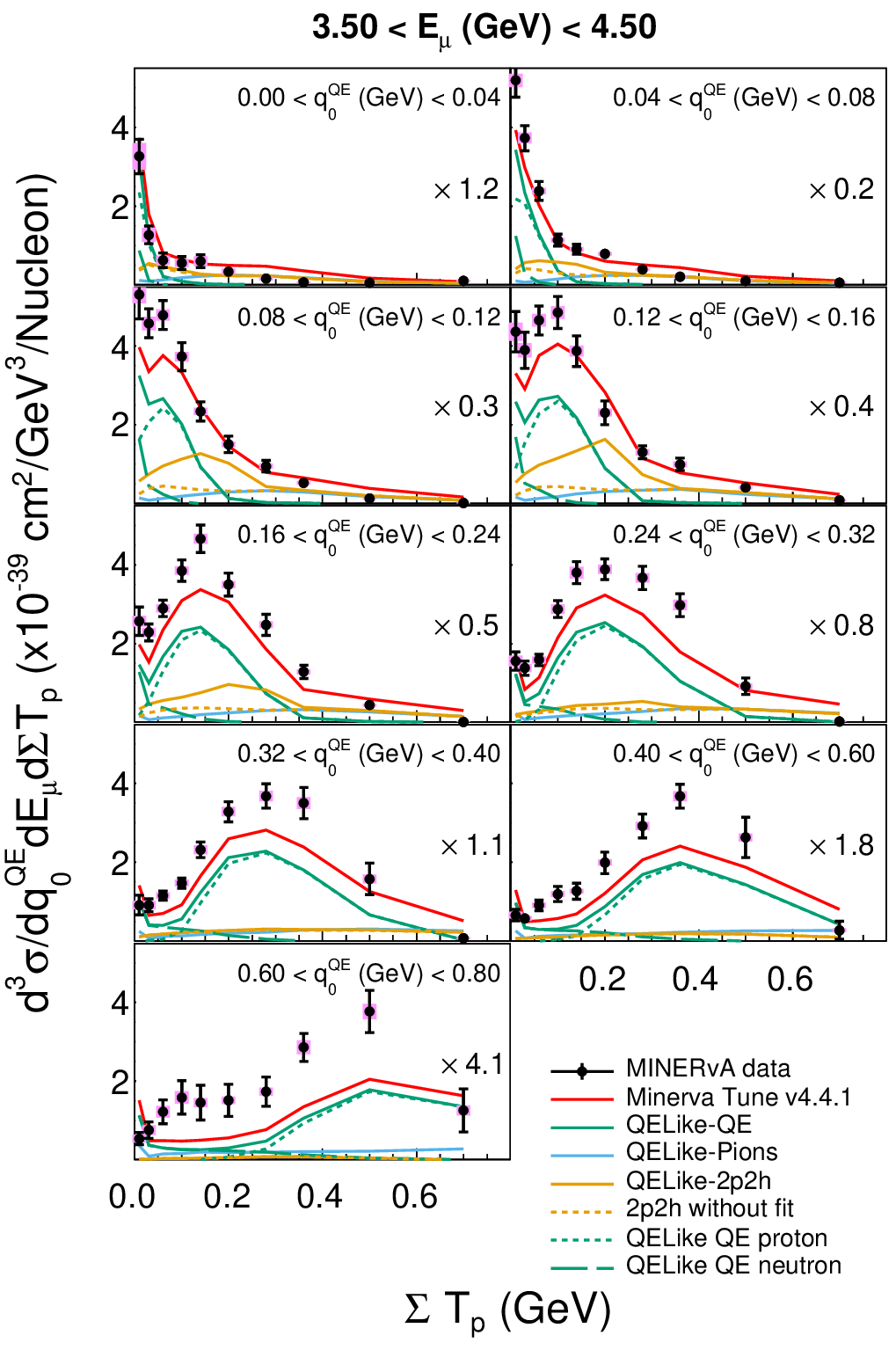}
    \caption{Flux-averaged triple differential cross section, $d^3\sigma/dE_\mu d\qzeroqe d\sumtp$.  Note that scale factors are applied to each of the $\qzeroqe$ panels.}
    \label{fig:sup_q0emusumtp_bin2}
\end{figure}

\begin{figure}[tp]
    \centering
    \includegraphics[width= 0.6\linewidth]{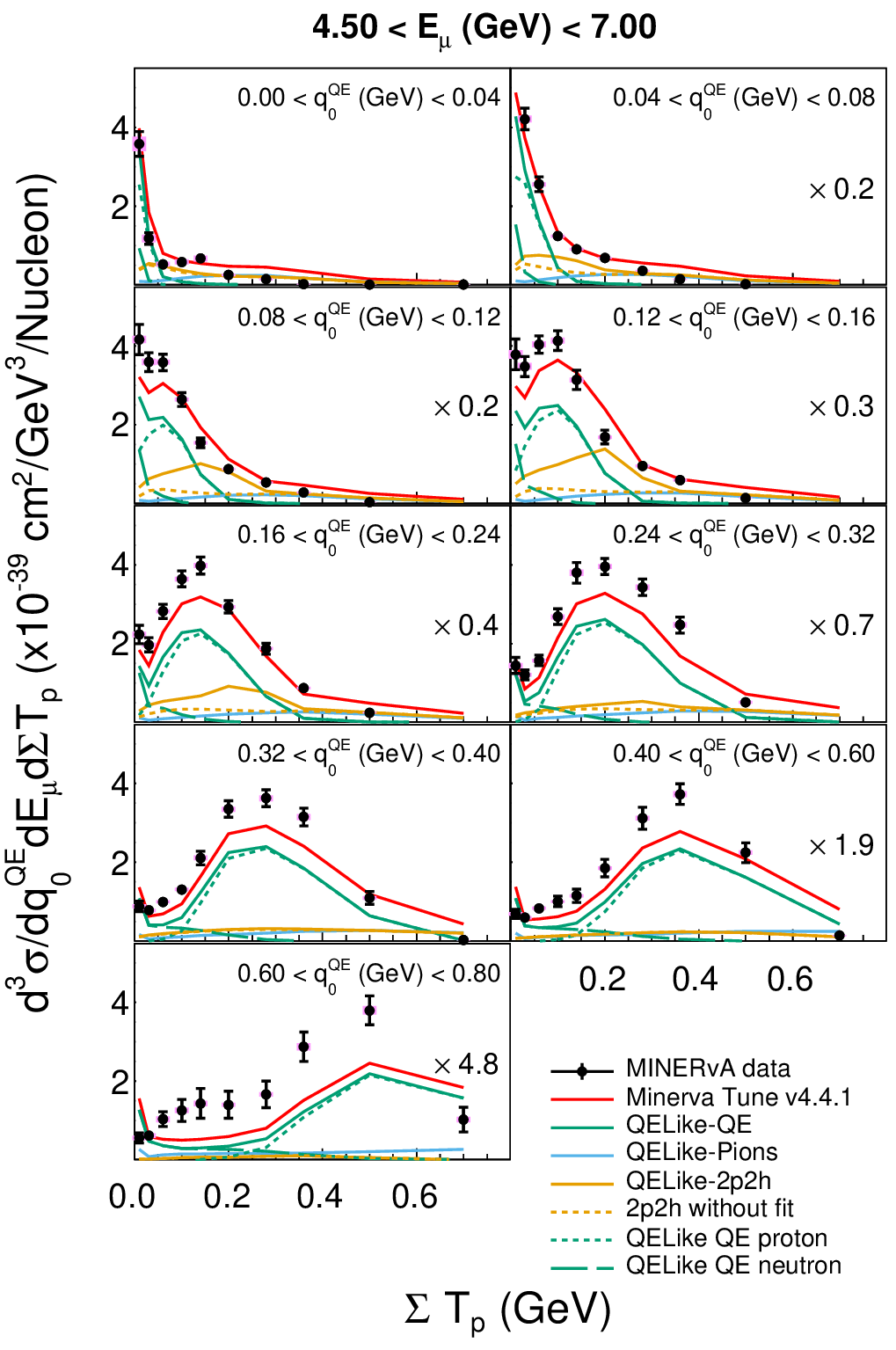}    \caption{Flux-averaged triple differential cross section, $d^3\sigma/dE_\mu d\qzeroqe d\sumtp$.  Note that scale factors are applied to each of the $\qzeroqe$ panels.}
    \label{fig:sup_q0emusumtp_bin3}
\end{figure}

\begin{figure}[tp]
    \centering
    \includegraphics[width= 0.6\linewidth]{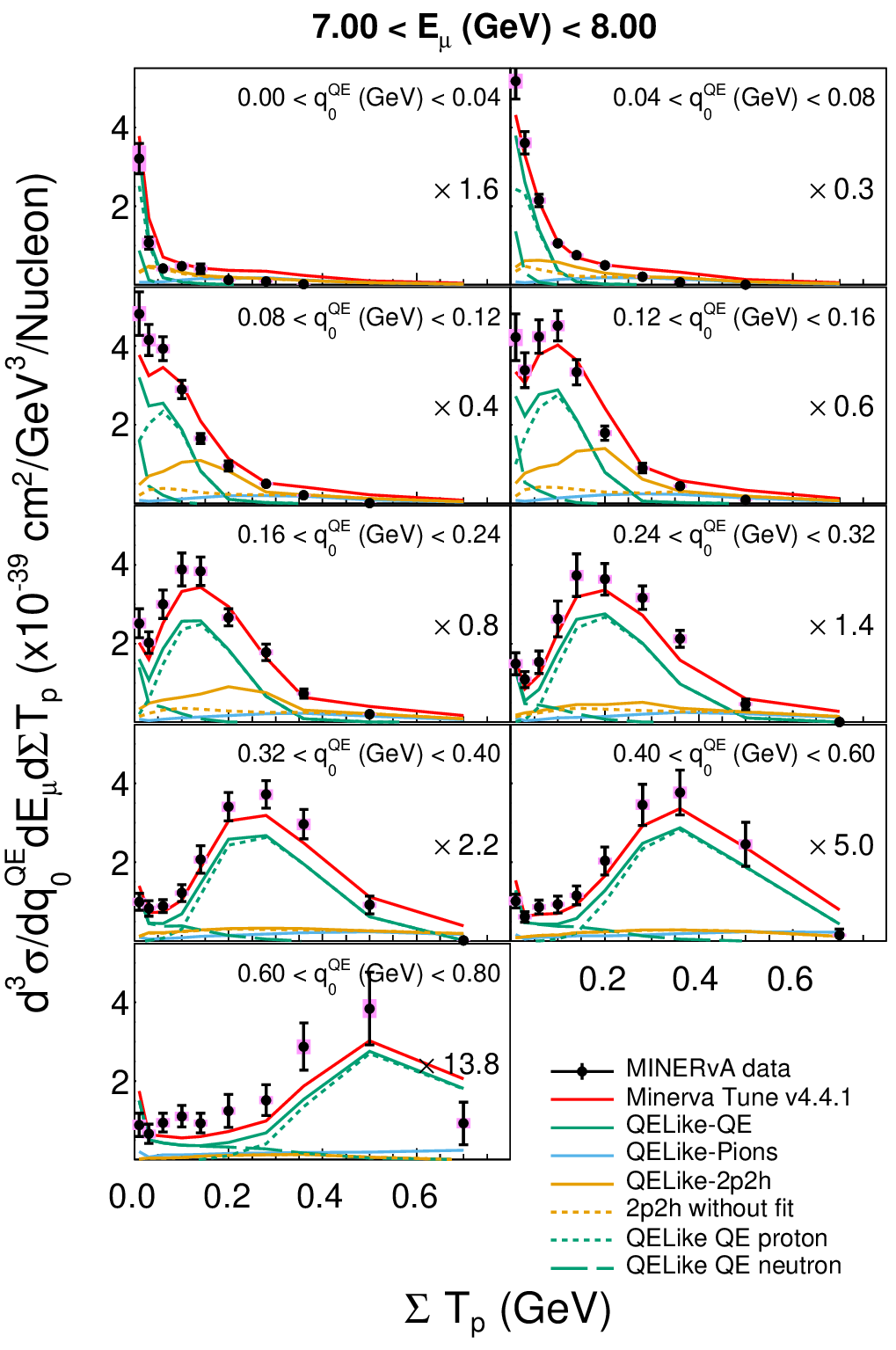}    \caption{Flux-averaged triple differential cross section, $d^3\sigma/dE_\mu d\qzeroqe d\sumtp$.  Note that scale factors are applied to each of the $\qzeroqe$ panels.}
    \label{fig:sup_q0emusumtp_bin4}
\end{figure}

\begin{figure}[tp]
    \centering
    \includegraphics[width= 0.6\linewidth]{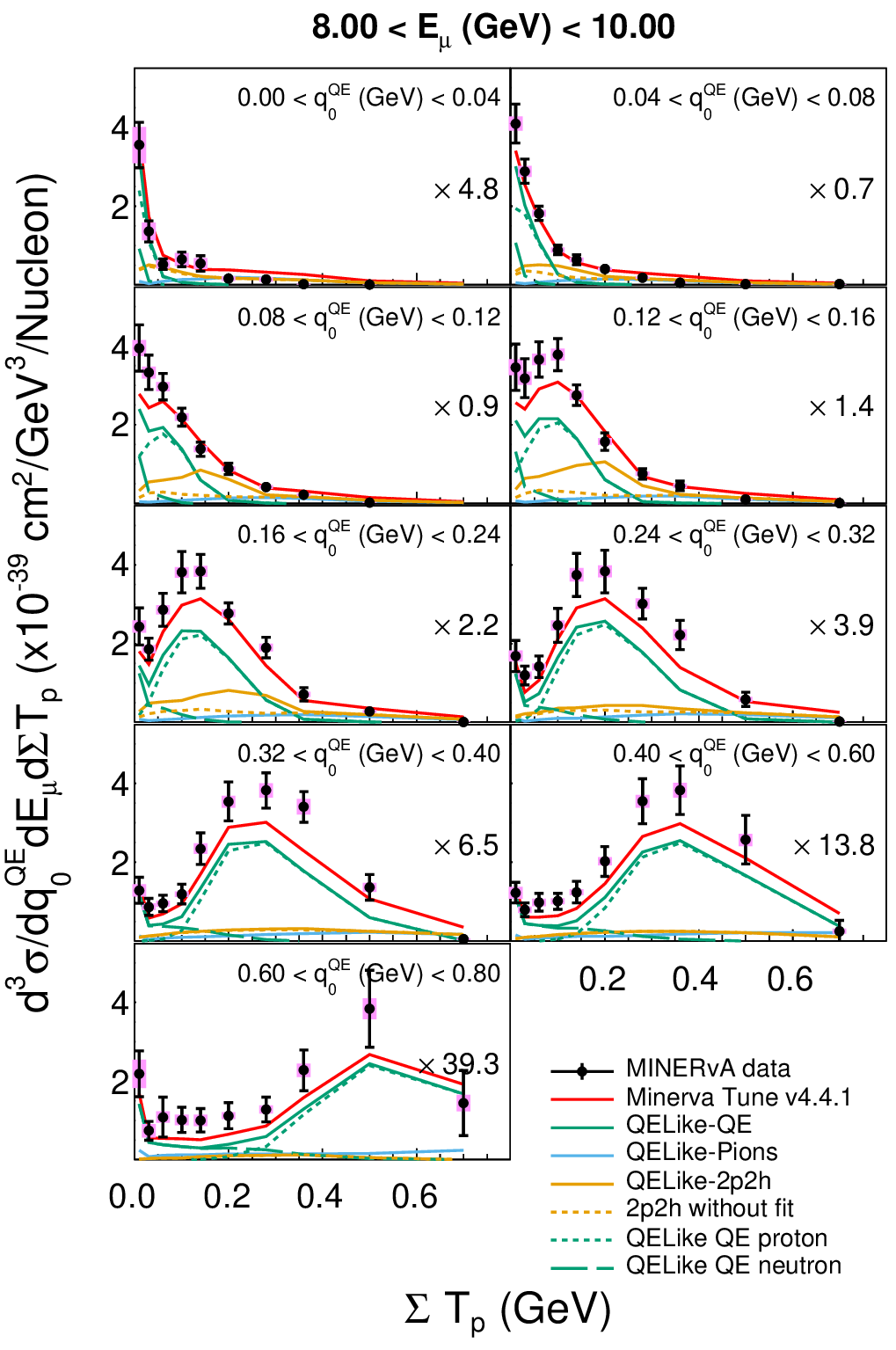}
    \caption{Flux-averaged triple differential cross section, $d^3\sigma/dE_\mu d\qzeroqe d\sumtp$.  Note that scale factors are applied to each of the $\qzeroqe$ panels.}
    \label{fig:sup_q0emusumtp_bin5}
\end{figure}

\begin{figure}[tp]
    \centering
    \includegraphics[width= 0.6\linewidth]{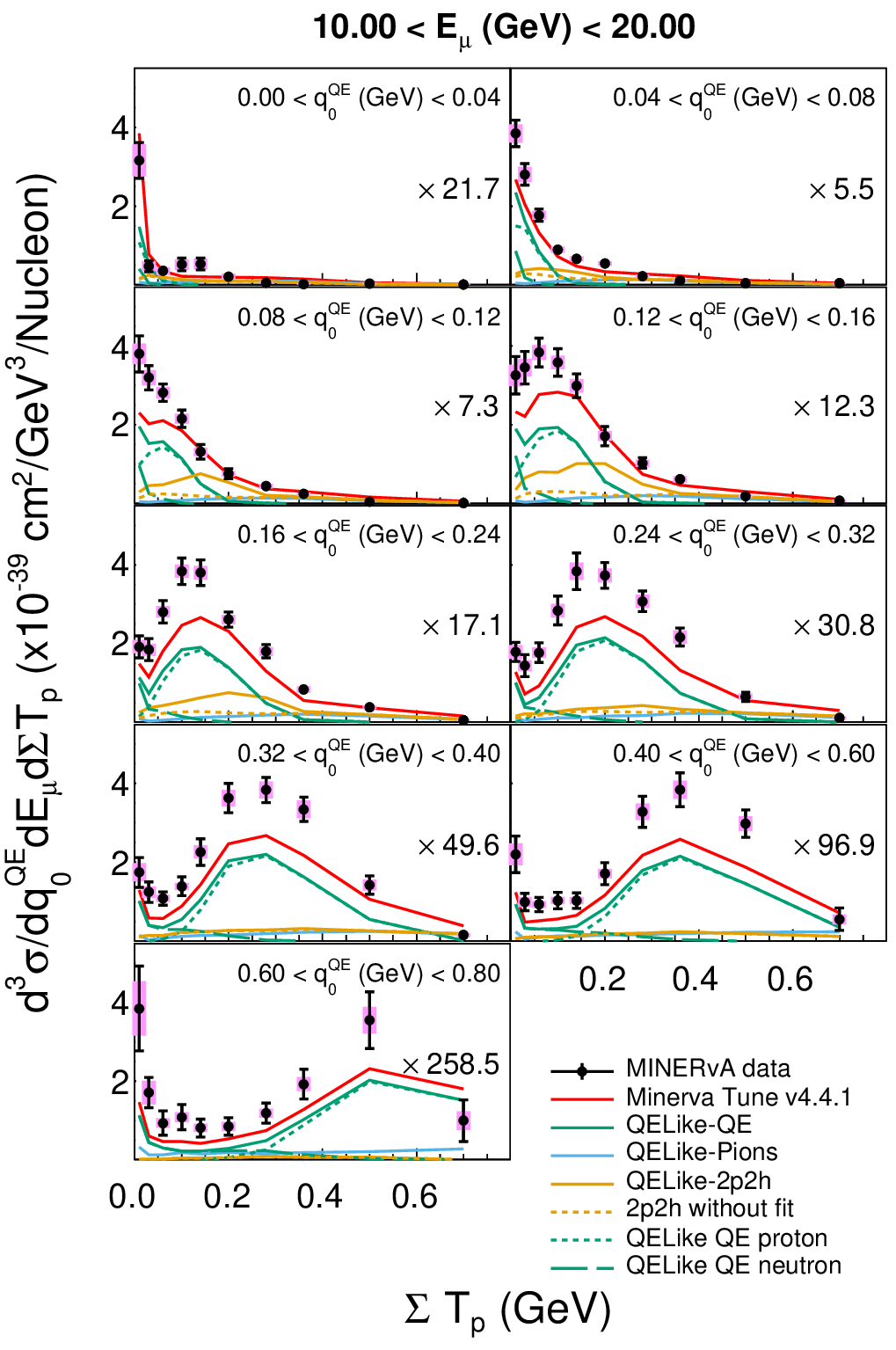}
    \caption{Flux-averaged triple differential cross section, $d^3\sigma/dE_\mu d\qzeroqe d\sumtp$.  Note that scale factors are applied to each of the $\qzeroqe$ panels.}
    \label{fig:sup_q0emusumtp_bin6}
\end{figure}

\begin{figure*}[tp]
    \centering
    \includegraphics[width= 0.9\linewidth]{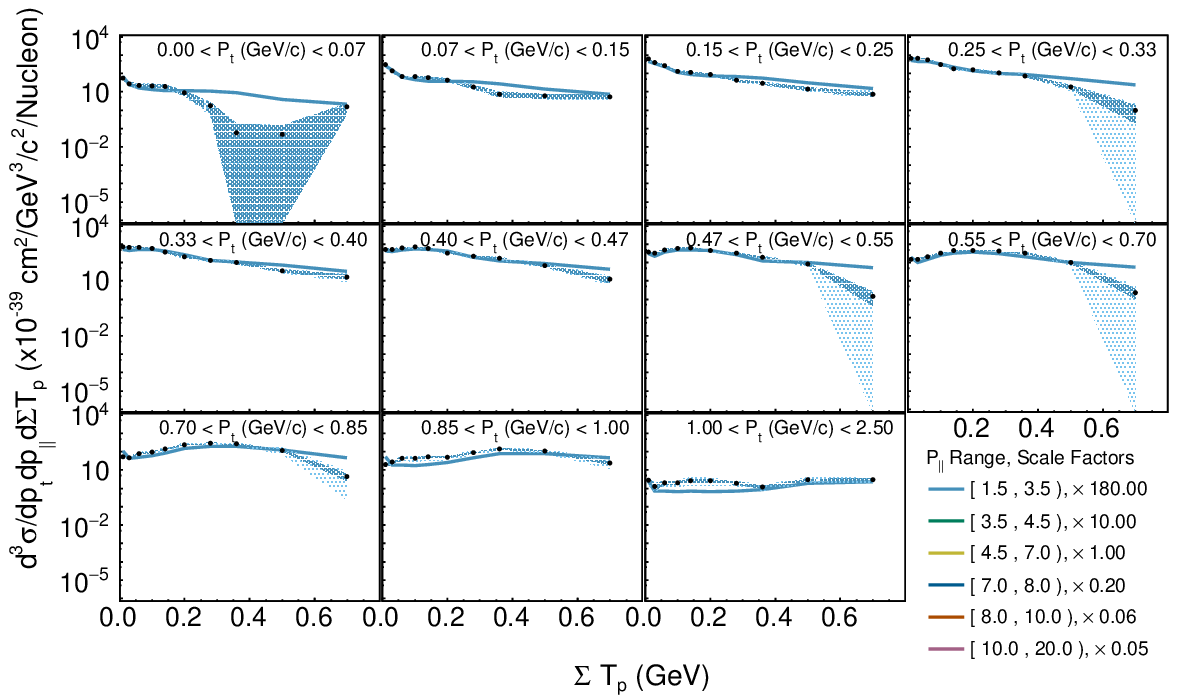}
    \caption{Triple differential cross section with only a single \pz~slice (1.5-3.5 GeV/c) shown The complete set of \pz~slices is shown in the main text, Figure 1.}
    \label{fig:sup_tripdiff_bin0}
\end{figure*}

\begin{figure*}[tp]
    \centering
    \includegraphics[width= 0.9\linewidth]{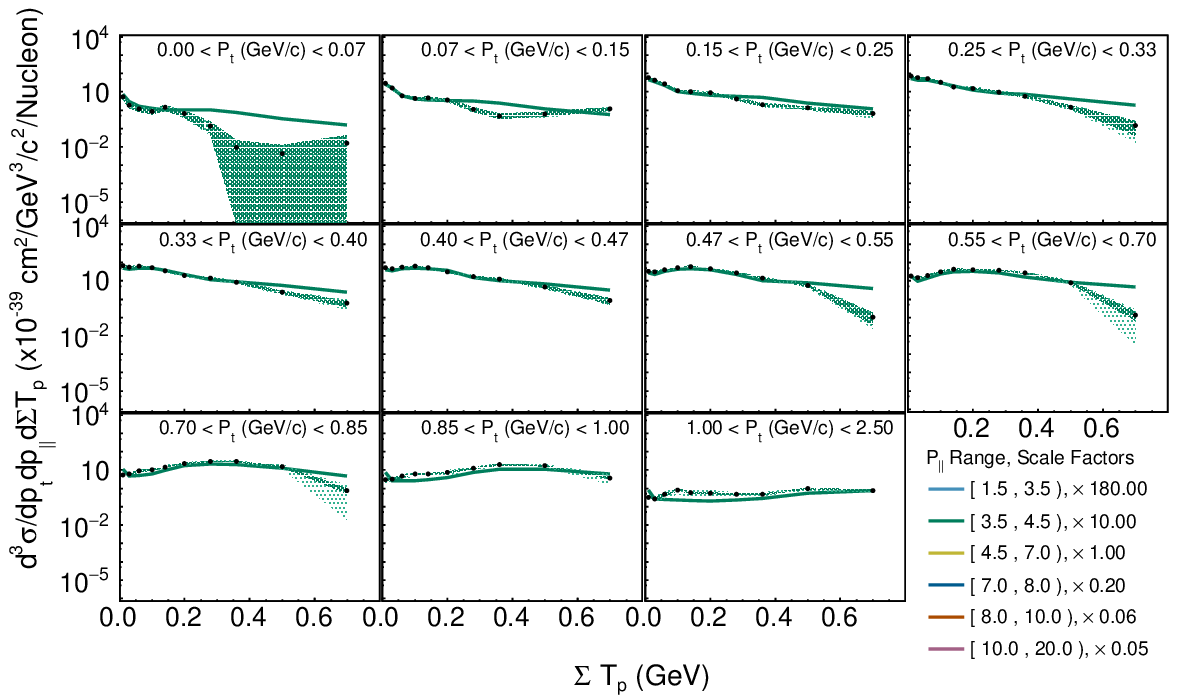}
    \caption{Triple differential cross section with only a single \pz~slice (3.5-4.5 GeV/c) shown The complete set of \pz~slices is shown in the main text, Figure 1.}
    \label{fig:sup_tripdiff_bin1}
\end{figure*}

\begin{figure*}[tp]
    \centering
    \includegraphics[width= 0.9\linewidth]{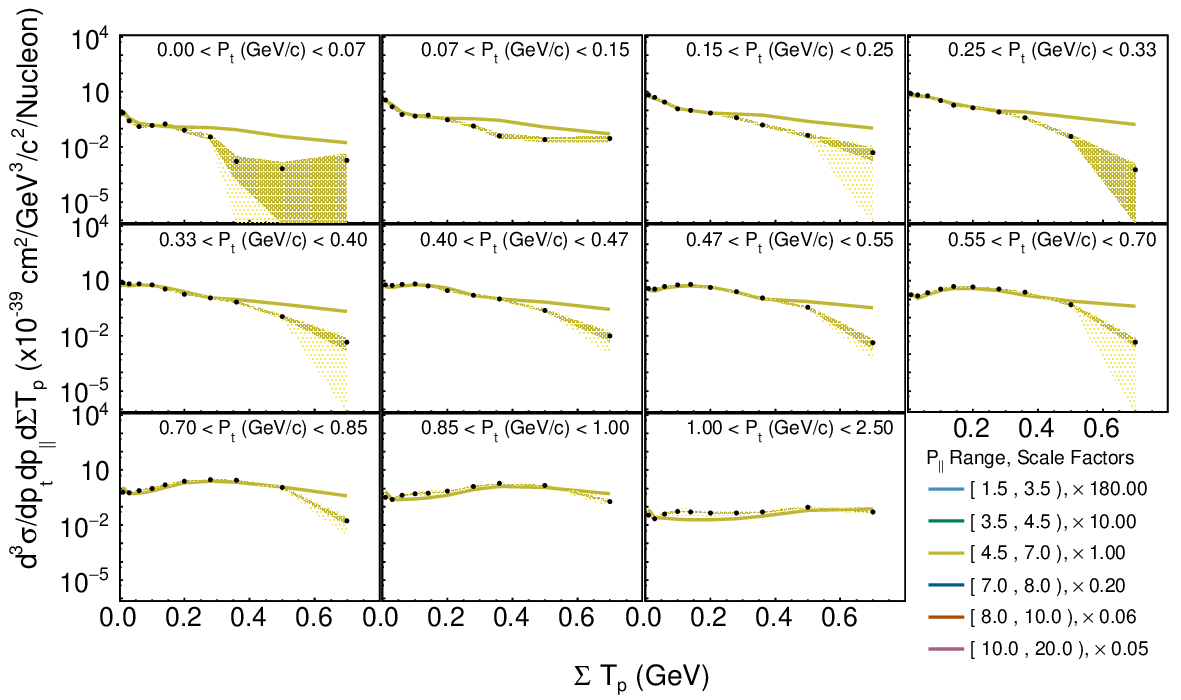}
    \caption{Triple differential cross section with only a single \pz~slice (4.5-7.0 GeV/c) shown The complete set of \pz~slices is shown in the main text, Figure 1.}
    \label{fig:sup_tripdiff_bin2}
\end{figure*}

\begin{figure*}[tp]
    \centering
    \includegraphics[width= 0.9\linewidth]{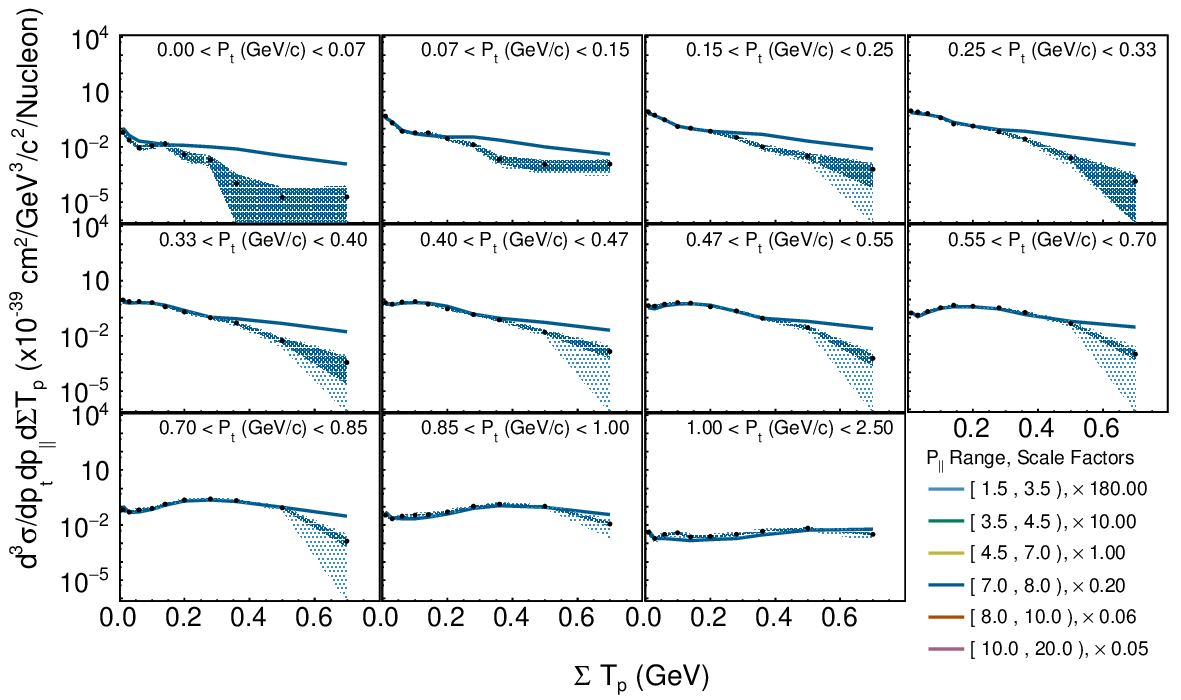}
    \caption{Triple differential cross section with only a single \pz~slice (7.0-8.0 GeV/c) shown The complete set of \pz~slices is shown in the main text, Figure 1.}
    \label{fig:sup_tripdiff_bin3}
\end{figure*}

\begin{figure*}[tp]
    \centering
    \includegraphics[width= 0.9\linewidth]{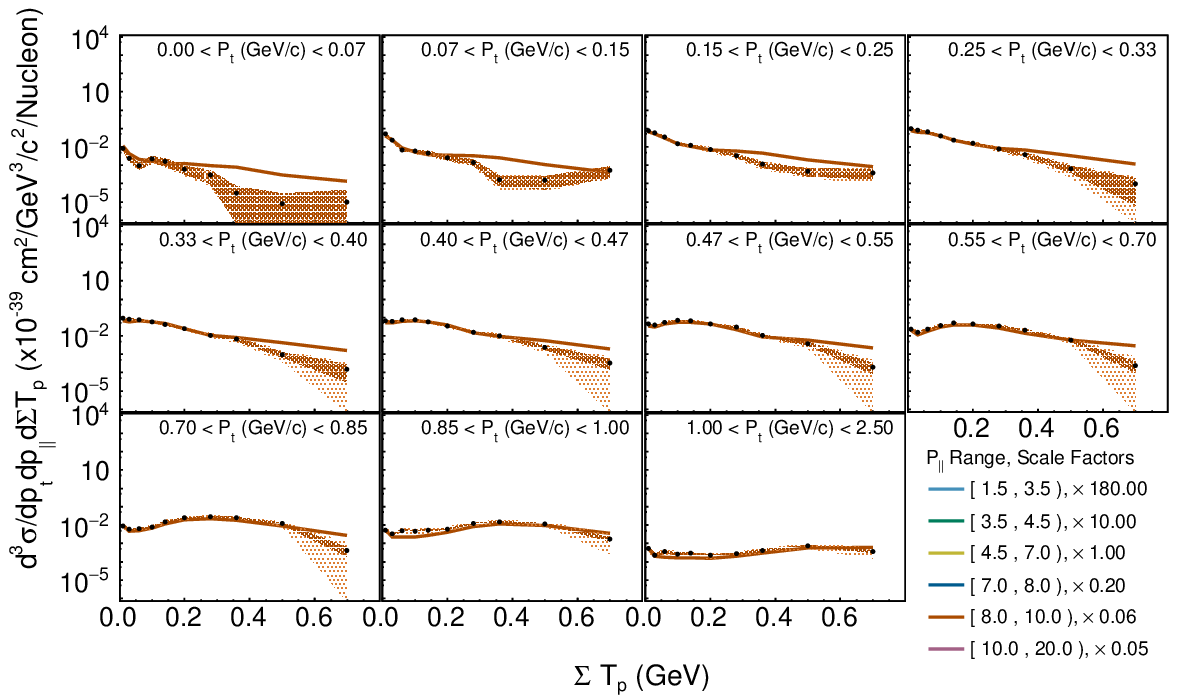}
    \caption{Triple differential cross section with only a single \pz~slice (8.0-10.0 GeV/c) shown The complete set of \pz~slices is shown in the main text, Figure 1.}
    \label{fig:sup_tripdiff_bin4}
\end{figure*}

Figures \ref{fig:sup_pzptsumtp_bin1}-\ref{fig:sup_pzptsumtp_bin6} show the ratio of the triple differential cross section to the reference model in the \pt-\pz-\sumtp, grouped in bins of $\pz$.
Figures \ref{fig:sup_q0emusumtp_bin1}-\ref{fig:sup_q0emusumtp_bin6} present the triple differential cross section in $\qzeroqe$-$E_\mu$-\sumtp, grouped in bins of $E_\mu$. Figures \ref{fig:sup_tripdiff_bin0}-\ref{fig:sup_tripdiff_bin4} split the main text's Figure 1 into individual \pz~slices to help disentangle regions with overlapping graphics.




\begin{figure*}[tp]
    \centering
    \includegraphics[width= 0.9\linewidth]{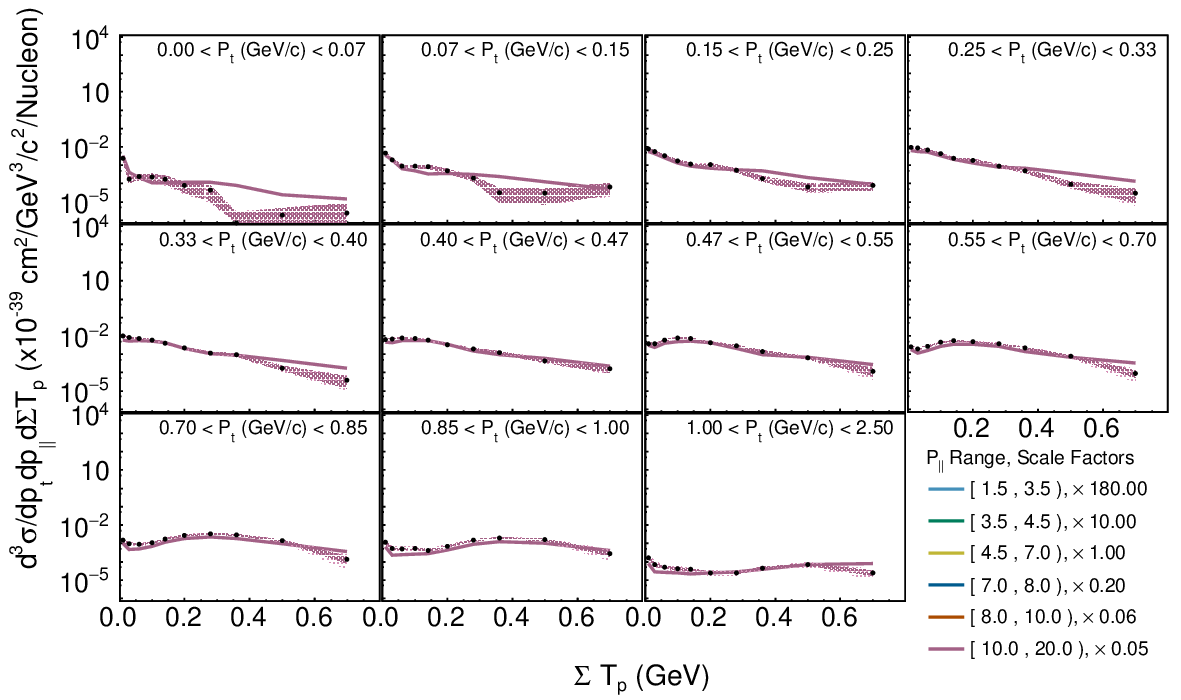}
    \caption{Triple differential cross section with only a single \pz~slice (10.0-20.0 GeV/c) shown The complete set of \pz~slices is shown in the main text, Figure 1.}
    \label{fig:sup_tripdiff_bin5}
\end{figure*}